\documentclass[final,5p,times,twocolumn,authoryear]{elsarticle}  
\usepackage{colortbl} 
\usepackage[normalem]{ulem}
\usepackage{amssymb}
\usepackage{graphicx}
\usepackage{txfonts}
\usepackage{array}
\journal{New Astronomy}
\usepackage[figuresright]{rotating}
\begin{document}
\begin{frontmatter}
  \title{Irregular emission cycles in the Oe star HD~60\,848\tnoteref{label1}}
  \tnotetext[label1]{Based on data collected with the NASA mission {\it Chandra}, and the ESA observatory {\it XMM-Newton}, an ESA Science Mission with instruments and contributions directly funded by ESA member states and the USA (NASA). Also based on optical spectra obtained with the TIGRE telescope, located at La Luz observatory, Mexico (TIGRE is a collaboration of the Hamburger Sternwarte, the Universities of Hamburg, Guanajuato and Li\`ege).}

\author[gr]{Gregor Rauw}
\ead{g.rauw@uliege.be}
\author[gr]{Ya\"el Naz\'e\fnref{fn1}}
\fntext[fn1]{Senior Research Associate FRS-FNRS (Belgium).}
\author[fc]{Fran Campos}
\author[jgf]{Joan Guarro Fl\'o}
\author[us]{Umberto Sollecchia}
  \address[gr]{Space sciences, Technologies and Astrophysics Research (STAR) Institute, Universit\'e de Li\`ege, All\'ee du 6 Ao\^ut, 19c, B\^at B5c, 4000 Li\`ege, Belgium}
  \address[fc]{Observatori Puig d'Agulles, Passatge Bosc 1, 08759 Vallirana, Barcelona, Spain}
  \address[jgf]{Observatori de Piera, Balmes, 2, 08784 Piera, Barcelona, Spain}   \address[us]{Via dei Malatesta 10, 67100 L'Aquila, Italy}
  \begin{abstract}
    We analyse the spectroscopic and photometric variability of the Oe star HD~60\,848 over the last twenty five years. The spectra reveal recurrent, but irregular cycles of increased circumstellar emission lines. These cycles are highly asymmetric displaying a slow increase over about 6 years, followed by a fast decay within about 6 months. Our analysis focuses on the most recent cycle (2013 -- 2020). The equivalent width and velocity separation of the emission peaks indicate variations of the outer disk radius by a factor $\sim 2.2$, although the variability appears more complex than expected from first principle relations for optically thin Keplerian disks. We observe a time delay between the variations of the strengths of He\,{\sc i} $\lambda$\,5876 on the one hand and H$\alpha$ and H$\beta$ on the other hand. We interpret this behaviour in a two-step disk growth scenario, where the disk first expands radially before its density increases. A difference in behaviour is also seen between H$\alpha$ and the H\,{\sc i} Paschen lines, with the latter displaying a more symmetric cycle, similar to the photometric variability. The rather fast decays of the H$\alpha$ emission observed in 2001, 2009 and 2018 -- 2019 suggest that the strong radiation field and early spectral type of the star lead to a faster dissipation of the disk than in later-type Be stars, as theoretically expected. We discuss X-ray observations of the star both during a high and a low-emission state. The X-ray spectrum is soft at both epochs, and the X-ray fluxes are only marginally different and remain consistent with the canonical $L_{\rm X}/L_{\rm bol}$ relation of O-type stars. These results indicate that the circumstellar decretion disk of HD~60\,848 has essentially no impact on the star's X-ray emission, and that the latter most likely arises inside the stellar wind.
  \end{abstract}

\end{frontmatter}
\section{Introduction}
About 20\% of the B-type stars in our Galaxy are classical Be stars, defined as non-supergiant B stars which display or have displayed H {\sc i} Balmer emission lines in their spectra \citep[for a review see][]{Riv13}. These emissions arise from circumstellar Keplerian decretion disks. Whilst Be stars are rapid rotators \citep{Riv13}, the exact role that rotation plays in the formation of the disk is not quite clear. Indeed, statistical studies indicate that Be stars rotate at less than critical velocity \citep{Zor16}. Oe stars form a much scarcer extension of the Be phenomenon into the O-star spectral range \citep{Con74,Neg04,Li18}. HD~60\,848 (= BN\,Gem = BD$+17^{\circ}$\,1623), the target of our present study, belongs to this latter category. The presence of hydrogen emission lines in the spectrum of HD~60\,848 is known for a long time \citep[see][and references therein]{Mer33}. The spectral type of the star was given as O5e in the Harvard classification system, but was revised to O8e by \citet{Pla24} and O7e by \citet{Moo32}. Accounting for the infilling of He\,{\sc i} lines by circumstellar emission, \citet{Neg04} proposed an O9.5\,IVe classification. 

The first evidence of variations of the spectrum of HD~60\,848 was obtained by \citet{Div83}. These authors presented a spectrum taken end of January 1983 where the H$\alpha$ emission line had an equivalent width (EW) of $-6.4$\,\AA. The line was significantly weaker than two years earlier (February 1981) when it displayed EW(H$\alpha$) = $-16.1$\,\AA\ \citep{And82}. The existence of emission line variations was confirmed by \citet{Rau07} who observed a similar transition between May 2001 and May 2002. A more extensive study was presented by \citet{Rau15}. In the latter paper, we notably pointed out a possible time lag in the variations of the lines formed in the circumstellar envelope of HD~60\,848. However, the sampling of the data analysed by \citet{Rau15} was rather scarce and heterogeneous. We thus set-up a more systematic monitoring using a modern echelle spectrograph as well as supporting data from amateur spectroscopists. 

Our optical campaign was supplemented by two X-ray observations. Among the Be stars, a small subset of objects, called the $\gamma$~Cas stars after their prototype, was found to display an unusually hard and bright thermal X-ray emission \citep{Smi16,Naz18}. Whilst most Be stars exhibit relatively soft X-ray spectra (with kT $\leq 1$\,keV) and luminosities below $10^{31}$\,erg\,s$^{-1}$, $\gamma$~Cas stars have X-ray plasma temperatures around 10\,keV and luminosities of $10^{32}$ -- $10^{33}$\,erg\,s$^{-1}$. The origin of the $\gamma$~Cas phenomenon is currently still debated. The most popular scenarios involve accretion either by a neutron star \citep[e.g.][]{Whi82,Pos17} or by a white dwarf \citep[e.g.][]{Ham16,Tsu18}, or magnetic interactions between the Be star and its disk \citep[e.g.][]{Smi98,Mot15}. Over recent years, one of these $\gamma$~Cas stars, the Oe star HD~45\,314, exhibited an episode of spectacular variations of its emission spectrum with transitions from a full Oe/Be state to a shell state and a nearly dissipation of the circumstellar disk \citep{Rau18}. During the near dissipation phase, the X-ray spectrum changed dramatically and the hard X-ray emission characteristic of the $\gamma$~Cas category faded considerably \citep{Rau18}. This observation pointed at the possibility that other Oe and Be stars might undergo similar transitions in their X-ray properties depending on the state of their circumstellar disk. In this context, we took advantage of the recent maximum emission state of HD~60\,848 to trigger a new X-ray observation to compare it with a previous one, obtained in April 2012 at a time when the star was in a low emission state \citep{Rau13}.

This paper reports the results of these two campaigns. In Sect.\,\ref{obser} we present the observational material, whereas Sect.\,\ref{results} describes our analyses of the optical and X-ray observations. The results of these investigations are discussed in Sect.\,\ref{Disc} and our conclusions are presented in Sect.\,\ref{Conclusion}.  
  
\section{Observations} \label{obser}
\subsection{Optical Spectroscopy}
Between autumn 2013 and spring 2020, we undertook a spectroscopic monitoring of HD~60\,848 with the fully robotic 1.2\,m TIGRE telescope \citep[formerly known as the Hamburg Robotic Telescope,][]{Schmitt} installed at La Luz Observatory near Guanajuato (Mexico). The TIGRE telescope features the refurbished HEROS echelle spectrograph covering the wavelength domain from 3800\,\AA\ to 8800\,\AA, with a small gap of 100\,\AA\ near 5800\,\AA. The resolving power is about 20\,000. The data reduction was performed with the dedicated TIGRE/HEROS reduction pipeline \citep{Mittag,Schmitt}. 

During the winter 2018--2019 observing season, the TIGRE data were complemented by observations from private amateur observatories taken by three of us. Spectra labelled JGF were obtained by co-author Joan Guarro Fl\'o using a 40.6\,cm Schmidt-Cassegrain telescope with a focal reducer to f/6.5. The telecope was equipped with a self-made echelle spectrograph providing a resolving power of 9000 with an ATIK\,460\,EX CCD camera binned $2 \times 2$. The data were taken either from Piera (Barcelona, Spain) or in remote operation from Santa Maria de Montmagastrell (Lleida, Spain). Spectra labelled FC were taken by Fran Campos using either a 35\,cm Ritchey-Chr\'etien telescope with f/8 or a 20\,cm Newton telescope with f/4.7 and a Barlow $2 \times$ lens resulting in an effective f/9.4. Both telescopes were installed in Vallirana (Barcelona, Spain). The spectrograph was a Baader DADOS with a 1200 l\,mm$^{-1}$ grating providing a resolving power of about 5000. Finally, data labelled US were collected by Umberto Sollecchia from L'Aquila (Italy) using a 23.5\,cm Celestron Schmidt-Cassegrain telescope. The spectrograph used a 1800 l\,mm$^{-1}$ grating offering a resolving power of 9000. All amateur data were processed with the ISIS software designed by C.\ Buil.

We further analysed a spectrum taken with the Aur\'elie spectrograph at the 1.52~m telescope at the Observatoire de Haute Provence in April 2006. This observation was taken with a 300 l\,mm$^{-1}$ grating providing a resolving power of 6000 over a 900\,\AA\ wide spectral domain centered at 8500\,\AA. Finally, we reconsidered all the data previously analysed in \citet{Rau15} and a few additional data were retrieved from the Be Star Spectra \citep[BeSS,][]{Nei11} database hosted at LESIA, Observatoire de Meudon, France \footnote{http://basebe.obspm.fr/basebe/}. 

\begin{figure}
    \begin{center}
      \resizebox{9cm}{!}{\includegraphics{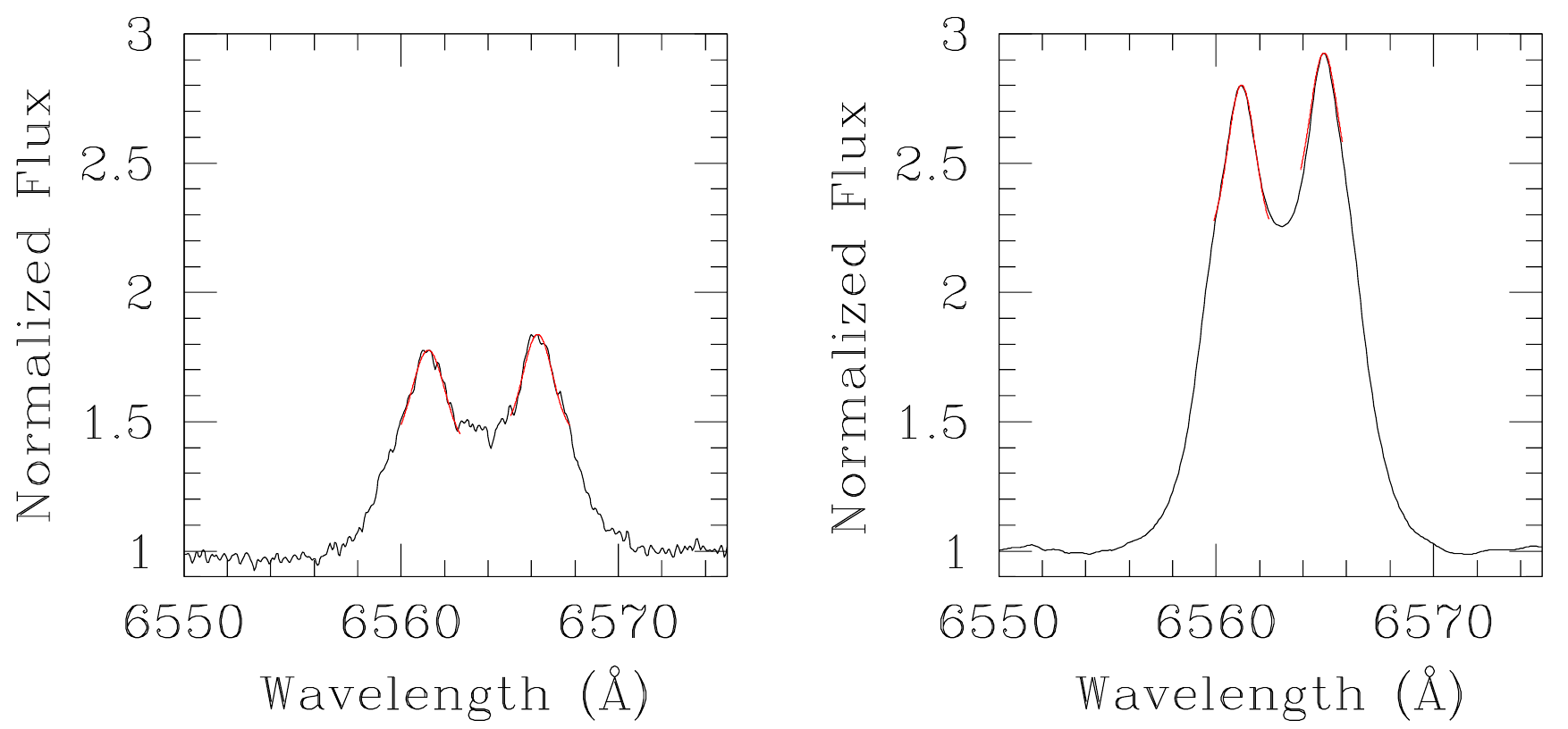}}
    \end{center}
\caption{Example of the fit of the peaks of the observed H$\alpha$ line profile (in black) with Gaussian functions (in red). The left panel illustrates the case of a relatively low-resolution, low S/N spectrum, whilst the right panel shows a high-resolution, high S/N spectrum.}
\label{gauss}
\end{figure}

For all the spectra we used the {\it telluric} tool within IRAF along with the atlas of telluric lines of \citet{Hinkle} to remove the telluric absorptions in the He\,{\sc i} $\lambda$\,5876 and H$\alpha$ regions. The spectra were continuum normalized using the MIDAS software adopting the same set of continuum windows for all data to achieve as homogeneous a normalization as possible. We measured the equivalent widths (EWs) of the H$\alpha$, H$\beta$ and He\,{\sc i} $\lambda$\,5876 emission lines by integrating the normalized spectra between 6550 and 6575\,\AA, 4850 and 4875\,\AA, and 5865 and 5885\,\AA, respectively. By fitting Gaussian profiles to the violet and red emission peaks of the profiles (see Fig.\,\ref{gauss}), we further determined the V/R ratio (defined as the ratio between the intensities above the continuum of the violet and red peak) as well as the velocity separation ($\Delta v$) between the peaks. The full set of H$\alpha$, H$\beta$ and He\,{\sc i} $\lambda$\,5876 measurements is listed in Table\,\ref{journal}. The quoted errors on the EWs are evaluated using the method of \citet{Vol06} which accounts for the S/N of the spectra and the actual line strength relative to the continuum. The errors on the V/R ratios depend mostly on the accuracy of the normalization and were found to be roughly equal to the inverse of the S/N ratio. Finally, the errors on $\Delta\,v$ depend upon the spectral resolution, the wavelength range included in the fit of the Gaussian profiles, and the line morphology (see Fig.\,\ref{gauss}). We have estimated these errors by varying the wavelength ranges over which the fit was performed and taking the dispersion of the resulting $\Delta\,v$ estimates.

\begin{table*}
  \caption{Journal of the spectroscopic observations of HD~60\,848 \label{journal}}
  \small
  \begin{tabular}{c r c c c c c c c c c}
    \hline
HJD-2\,400\,000 &  \multicolumn{3}{c}{H$\alpha$} & & \multicolumn{2}{c}{H$\beta$} & &\multicolumn{2}{c}{He\,{\sc i} $\lambda$\,5876} & Instrument \\
\cline{2-4}\cline{6-7}\cline{9-10}
                &  EW & V/R & $\Delta\,v$ & & EW & $\Delta\,v$ & & EW & $\Delta\,v$ \\ 
& (\AA) & & (km\,s$^{-1}$) & & (\AA) & (km\,s$^{-1}$) & & (\AA) & (km\,s$^{-1}$) \\
\hline
49713.132 &         &      &       &&          &       && $ -1.17 \pm 0.24$& $296.3 \pm 7.0$ & AAT \\
49713.134 &         &      &       && $ -0.08 \pm 0.20$& $249.4 \pm 7.0$  &&          &       & AAT \\
50829.713 &         &      &       && $ -0.63 \pm 0.20$& $229.7 \pm 1.8$ &&          &       & INT \\
52190.764 &         &      &       && $ -0.14 \pm 0.45$& $228.5 \pm 6.2$ &&          &       & INT \\
52301.626 & $ -5.5 \pm 0.1$ & $1.03 \pm 0.01$& $174.5 \pm 1.3$ &&          &       &&          &       & INT \\
50506.444 & $ -7.3 \pm 0.4$ & $1.01 \pm 0.01$& $200.2 \pm 2.2$ &&          &       &&          &       & Aur\'elie \\
50507.440 & $ -7.7 \pm 0.3$ & $1.02 \pm 0.01$ & $201.4 \pm 2.2$ &&          &       &&          &       & Aur\'elie \\
50508.433 & $ -7.2 \pm 0.4$ & $1.03 \pm 0.01$ & $202.7 \pm 2.2$ &&          &       &&          &       & Aur\'elie \\
50508.451 & $ -7.2 \pm 0.4$ & $1.01 \pm 0.01$ & $202.3 \pm 2.2$ &&          &       &&          &       & Aur\'elie \\
50510.353 & $ -7.2 \pm 0.3$ & $0.99 \pm 0.01$ & $200.7 \pm 2.2$ &&          &       &&          &       & Aur\'elie \\
50510.369 & $ -6.8 \pm 0.4$ & $0.97 \pm 0.01$ & $200.4 \pm 2.2$ &&          &       &&          &       & Aur\'elie \\
51132.531 & $ -7.3 \pm 0.1$ & $1.01 \pm 0.01$ & $171.2 \pm 2.2$ &&          &       &&          &       & Aur\'elie \\
51133.531 & $ -7.3 \pm 0.2$ & $1.00 \pm 0.01$ & $164.7 \pm 2.2$ &&          &       &&          &       & Aur\'elie \\
51134.684 & $ -7.1 \pm 0.2$ & $1.00 \pm 0.01$ & $167.7 \pm 2.2$ &&          &       &&          &       & Aur\'elie \\
51135.570 & $ -7.2 \pm 0.2$ & $1.04 \pm 0.01$ & $168.9 \pm 2.2$ &&          &       &&          &       & Aur\'elie \\
51136.558 & $ -7.1 \pm 0.2$ & $1.02 \pm 0.01$ & $169.7 \pm 2.2$ &&          &       &&          &       & Aur\'elie \\
51136.698 &         &      &       && $ -0.87 \pm 0.15$& $207.9 \pm 6.5$&&          &       & Aur\'elie \\
51137.555 & $ -7.4 \pm 0.2$& $1.00 \pm 0.01$ & $165.9 \pm 2.2$ &&          &       &&          &       & Aur\'elie \\
51819.663 &         &      &       && $ -1.40 \pm 0.07$& $199.7 \pm 6.5$ &&          &       & Aur\'elie \\
51821.660 &         &      &       && $ -1.42 \pm 0.10$& $202.7 \pm 6.5$&&          &       & Aur\'elie \\
54472.613 &         &      &       &&          &       && $ -1.51 \pm 0.10$& $250.9 \pm 6.5$ & Aur\'elie \\
54473.344 & $ -11.5 \pm 0.4$& $1.06 \pm 0.01$ & $172.8 \pm 2.2$ &&          &       &&          &       & Aur\'elie \\
54474.605 &         &      &       &&          &       && $ -1.45 \pm 0.20$& $250.5 \pm 6.5$ & Aur\'elie \\
54475.599 &         &      &       &&          &       && $ -1.52 \pm 0.10$& $255.1 \pm 6.5$ & Aur\'elie \\
51299.481 & $ -9.0 \pm 0.3$& $0.97 \pm 0.01$ & $161.7 \pm 2.5$ && $ -1.06 \pm 0.18$& $203.7 \pm 5.2$&& $ -0.40 \pm 0.15$& $253.8 \pm 4.0$ & FEROS \\
51299.489 & $ -8.9 \pm 0.3$& $0.98 \pm 0.01$ & $161.9 \pm 2.5$ && $ -1.07 \pm 0.19$& $204.9 \pm 5.2$ && $ -0.35 \pm 0.09$& $253.5 \pm 4.0$ & FEROS \\
51300.482 & $ -8.8 \pm 0.3$& $0.95 \pm 0.01$ & $163.2 \pm 2.5$ && $ -1.07 \pm 0.20$& $204.8 \pm 5.2$ && $ -0.34 \pm 0.13$& $251.0 \pm 4.0$ & FEROS \\
51300.488 & $ -8.8 \pm 0.2$& $0.95 \pm 0.01$ & $166.8 \pm 2.5$ && $ -1.05 \pm 0.20$& $205.8 \pm 5.2$ && $ -0.42 \pm 0.14$& $253.6 \pm 4.0$ & FEROS \\
51301.477 & $ -8.9 \pm 0.2$& $0.97 \pm 0.01$ & $160.4 \pm 2.5$ && $ -1.08 \pm 0.22$& $206.9 \pm 5.2$ && $ -0.33 \pm 0.11$& $255.4 \pm 4.0$ & FEROS \\
51301.483 & $ -8.9 \pm 0.4$& $0.97 \pm 0.01$ & $158.0 \pm 2.5$ && $ -0.87 \pm 0.36$& $205.9 \pm 5.2$ && $ -0.33 \pm 0.17$& $248.9 \pm 4.0$ & FEROS \\
51302.481 & $ -9.0 \pm 0.4$& $0.95 \pm 0.01$ & $165.3 \pm 2.5$ && $ -1.18 \pm 0.24$& $214.7 \pm 5.2$ && $ -0.28 \pm 0.18$& $259.5 \pm 4.0$ & FEROS \\
51302.487 & $ -9.0 \pm 0.3$& $0.96 \pm 0.01$ & $159.8 \pm 2.5$ && $ -1.17 \pm 0.22$& $215.2 \pm 5.2$ && $ -0.34 \pm 0.14$& $264.1 \pm 4.0$ & FEROS \\
51304.469 & $ -8.9 \pm 0.3$& $0.97 \pm 0.01$ & $154.9 \pm 2.5$ && $ -1.15 \pm 0.25$& $215.9 \pm 5.2$ && $ -0.21 \pm 0.17$& $260.5 \pm 4.0$ & FEROS \\
51304.475 & $ -8.9 \pm 0.4$& $0.97 \pm 0.01$ & $153.9 \pm 2.5$ && $ -1.12 \pm 0.26$& $212.5 \pm 5.2$ && $ -0.24 \pm 0.20$& $253.9 \pm 4.0$ & FEROS \\
51669.463 & $ -13.8 \pm 0.3$& $1.06 \pm 0.01$ & $149.2 \pm 2.5$ && $ -1.33 \pm 0.32$& $197.7 \pm 5.2$ && $ -1.36 \pm 0.18$& $231.5 \pm 4.0$ & FEROS \\
51669.469 & $ -13.9 \pm 0.4$& $1.08 \pm 0.01$ & $141.9 \pm 2.5$ && $ -1.33 \pm 0.44$& $196.3 \pm 5.2$ && $ -1.74 \pm 0.27$& $229.6 \pm 4.0$ & FEROS \\
51671.456 & $ -12.3 \pm 0.5$& $1.07 \pm 0.01$ & $152.1 \pm 2.5$ &&          &       &&          &       & FEROS \\
51671.461 & $ -12.7 \pm 0.4$& $1.06 \pm 0.01$ & $153.3 \pm 2.5$ &&          &       &&          &       & FEROS \\
51673.454 & $ -13.8 \pm 0.3$& $1.10 \pm 0.01$ & $148.5 \pm 2.5$ && $ -1.49 \pm 0.23$& $194.2 \pm 5.2$ && $ -1.33 \pm 0.23$& $231.4 \pm 4.0$ & FEROS \\
51673.463 & $ -13.7 \pm 0.3$& $1.10 \pm 0.01$ & $141.9 \pm 2.5$ && $ -1.49 \pm 0.20$& $195.7 \pm 5.2$ && $ -1.34 \pm 0.12$& $232.7 \pm 4.0$ & FEROS \\
52039.463 & $ -14.1 \pm 0.5$& $1.01 \pm 0.02$ & $141.9 \pm 2.5$ && $ -1.36 \pm 0.62$& $189.8 \pm 5.2$ && $ -1.81 \pm 0.54$& $217.5 \pm 4.0$ & FEROS \\
52039.471 & $ -14.0 \pm 0.4$& $1.02 \pm 0.01$ & $152.2 \pm 2.5$ && $ -1.42 \pm 0.32$& $186.2 \pm 5.2$ && $ -1.88 \pm 0.15$& $216.4 \pm 4.0$ & FEROS \\
52039.513 & $ -14.2 \pm 0.4$& $1.02 \pm 0.01$ & $147.2 \pm 2.5$ && $ -1.41 \pm 0.44$& $188.6 \pm 5.2$ && $ -2.04 \pm 0.26$& $215.8 \pm 4.0$ & FEROS \\
52335.554 & $ -5.7 \pm 0.2$& $1.01 \pm 0.01$ & $211.1 \pm 2.5$ && $ +0.31 \pm 0.18$& $277.2 \pm 5.2$ && $ -0.66 \pm 0.15$& $332.9 \pm 4.0$ & FEROS \\
52337.540 & $ -5.8 \pm 0.3$& $1.15 \pm 0.01$ & $206.7 \pm 2.5$ && $ +0.27 \pm 0.18$& $275.3 \pm 5.2$ && $ -0.67 \pm 0.12$& $327.6 \pm 4.0$ & FEROS \\
52339.524 & $ -6.2 \pm 0.2$& $0.97 \pm 0.01$ & $211.5 \pm 2.5$ && $ +0.21 \pm 0.14$& $275.4 \pm 5.2$ && $ -0.73 \pm 0.07$& $322.3 \pm 4.0$ & FEROS \\
51939.775 & $ -15.4 \pm 0.5$& $0.96 \pm 0.01$ & $158.8 \pm 1.6$ && $ -1.56 \pm 0.45$& $205.3 \pm 8.5$&& $ -1.75 \pm 0.19$& $239.9 \pm 4.5$ & UVES \\
53331.175 & $ -6.9 \pm 0.2$& $1.01 \pm 0.01$ & $198.6 \pm 3.4$ && $ -0.70 \pm 0.26$& $241.7 \pm 2.0$&& $ -0.49 \pm 0.25$& $280.7 \pm 3.0$ & Elodie \\
54818.788 & $ -11.6 \pm 0.2$& $1.01 \pm 0.01$ & $152.8 \pm 10.0$ &&          &       &&          &       & Mons \\
54818.803 & $ -10.6 \pm 0.3$& $1.01 \pm 0.01$ & $155.7 \pm 10.0$ &&          &       &&          &       & Mons \\
54822.701 & $ -10.6 \pm 0.3$& $1.03 \pm 0.01$ & $144.7 \pm 10.0$ &&          &       &&          &       & Mons \\
54822.715 & $ -10.7 \pm 0.3$& $1.03 \pm 0.01$ & $144.2 \pm 10.0$ &&          &       &&          &       & Mons \\
54822.729 & $ -11.0 \pm 0.2$& $1.02 \pm 0.01$ & $133.3 \pm 10.0$ &&          &       &&          &       & Mons \\
54822.743 & $ -10.7 \pm 0.2$& $1.04 \pm 0.01$ & $140.2 \pm 10.0$ &&          &       &&          &       & Mons \\
54829.663 & $ -10.1 \pm 0.3$& $1.05 \pm 0.01$ & $144.9 \pm 10.0$ &&          &       &&          &       & Mons \\
54835.730 & $ -10.3 \pm 0.2$& $0.98 \pm 0.01$ & $153.2 \pm 10.0$ &&          &       &&          &       & Mons \\
54835.744 & $ -10.3 \pm 0.1$& $0.98 \pm 0.01$ & $151.9 \pm 10.0$ &&          &       &&          &       & Mons \\
\hline
  \end{tabular}
\end{table*}
\addtocounter{table}{-1}
\begin{table*}
  \caption{Continued}
  \small
  \begin{tabular}{c r c c c c c c c c c}
    \hline
HJD-2\,400\,000 &  \multicolumn{3}{c}{H$\alpha$} & & \multicolumn{2}{c}{H$\beta$} & &\multicolumn{2}{c}{He\,{\sc i} $\lambda$\,5876} & Instrument \\
\cline{2-4}\cline{6-7}\cline{9-10}
                &  EW & V/R & $\Delta\,v$ & & EW & $\Delta\,v$ & & EW & $\Delta\,v$ \\ 
& (\AA) & & (km\,s$^{-1}$) & & (\AA) & (km\,s$^{-1}$) & & (\AA) & (km\,s$^{-1}$) \\
\hline
54835.758 & $ -10.4 \pm 0.2$& $0.98 \pm 0.01$ & $153.8 \pm 10.0$ &&          &       &&          &       & Mons \\
54850.710 & $ -10.3 \pm 0.3$& $1.01 \pm 0.01$ & $150.0 \pm 10.0$ &&          &       &&          &       & Mons \\
54850.724 & $ -9.7 \pm 0.2$& $1.00 \pm 0.01$ & $156.0 \pm 10.0$ &&          &       &&          &       & Mons \\
54850.738 & $ -10.4 \pm 0.3$& $1.01 \pm 0.01$ & $142.9 \pm 10.0$ &&          &       &&          &       & Mons \\
54852.702 & $ -9.8 \pm 0.2$& $0.99 \pm 0.01$ & $148.9 \pm 10.0$ &&          &       &&          &       & Mons \\
54852.716 & $ -10.9 \pm 0.3$& $0.99 \pm 0.01$ & $144.5 \pm 10.0$ &&          &       &&          &       & Mons \\
54852.730 & $ -10.5 \pm 0.3$& $0.98 \pm 0.01$ & $144.4 \pm 10.0$ &&          &       &&          &       & Mons \\
54853.704 & $ -9.9 \pm 0.3$& $0.95 \pm 0.01$ & $149.5 \pm 10.0$ &&          &       &&          &       & Mons \\
54853.718 & $ -10.2 \pm 0.2$& $0.95 \pm 0.01$ & $150.6 \pm 10.0$ &&          &       &&          &       & Mons \\
54853.732 & $ -10.3 \pm 0.4$& $0.95 \pm 0.01$ & $152.1 \pm 10.0$ &&          &       &&          &       & Mons \\
54854.701 & $ -9.4 \pm 0.1$& $0.95 \pm 0.01$ & $149.8 \pm 10.0$ &&          &       &&          &       & Mons \\
54854.715 & $ -9.5 \pm 0.2$& $0.94 \pm 0.01$ & $149.0 \pm 10.0$ &&          &       &&          &       & Mons \\
54854.729 & $ -9.5 \pm 0.2$& $0.94 \pm 0.01$ & $148.9 \pm 10.0$ &&          &       &&          &       & Mons \\
54858.657 & $ -9.6 \pm 0.2$& $1.09 \pm 0.01$ & $137.8 \pm 10.0$ &&          &       &&          &       & Mons \\
54858.671 & $ -9.4 \pm 0.2$& $1.09 \pm 0.01$ & $137.4 \pm 10.0$ &&          &       &&          &       & Mons \\
54858.685 & $ -9.0 \pm 0.2$& $1.09 \pm 0.01$ & $131.6 \pm 10.0$ &&          &       &&          &       & Mons \\
54859.689 & $ -10.1 \pm 0.2$& $1.10 \pm 0.01$ & $128.9 \pm 10.0$ &&          &       &&          &       & Mons \\
54859.703 & $ -8.9 \pm 0.3$& $1.12 \pm 0.01$ & $136.0 \pm 10.0$ &&          &       &&          &       & Mons \\
54859.717 & $ -9.8 \pm 0.3$& $1.10 \pm 0.01$ & $124.6 \pm 10.0$ &&          &       &&          &       & Mons \\
54860.656 & $ -9.3 \pm 0.2$& $1.11 \pm 0.01$ & $133.7 \pm 10.0$ &&          &       &&          &       & Mons \\
54860.670 & $ -9.7 \pm 0.3$& $1.10 \pm 0.01$ & $136.8 \pm 10.0$ &&          &       &&          &       & Mons \\
54860.684 & $ -9.3 \pm 0.2$& $1.11 \pm 0.01$ & $145.2 \pm 10.0$ &&          &       &&          &       & Mons \\
54861.632 & $ -9.4 \pm 0.3$& $1.06 \pm 0.01$ & $142.4 \pm 10.0$ &&          &       &&          &       & Mons \\
54861.646 & $ -9.3 \pm 0.3$& $1.08 \pm 0.01$ & $152.9 \pm 10.0$ &&          &       &&          &       & Mons \\
54865.658 & $ -8.9 \pm 0.2$& $0.95 \pm 0.01$ & $149.7 \pm 10.0$ &&          &       &&          &       & Mons \\
54865.672 & $ -8.7 \pm 0.2$& $0.95 \pm 0.01$ & $150.1 \pm 10.0$ &&          &       &&          &       & Mons \\
54865.686 & $ -9.3 \pm 0.4$& $0.94 \pm 0.01$ & $139.8 \pm 10.0$ &&          &       &&          &       & Mons \\
54871.634 & $ -8.7 \pm 0.5$& $1.02 \pm 0.01$ & $158.5 \pm 10.0$ &&          &       &&          &       & Mons \\
54871.648 & $ -8.5 \pm 0.3$& $1.00 \pm 0.01$ & $152.9 \pm 10.0$ &&          &       &&          &       & Mons \\
54873.671 & $ -7.5 \pm 0.3$& $0.93 \pm 0.01$ & $151.5 \pm 10.0$ &&          &       &&          &       & Mons \\
54873.691 & $ -8.0 \pm 0.3$& $0.94 \pm 0.01$ & $156.0 \pm 10.0$ &&          &       &&          &       & Mons \\
54873.710 & $ -8.4 \pm 0.4$& $0.93 \pm 0.01$ & $140.0 \pm 10.0$ &&          &       &&          &       & Mons \\
54874.671 & $ -8.4 \pm 0.2$& $0.97 \pm 0.01$ & $121.9 \pm 10.0$ &&          &       &&          &       & Mons \\
54879.704 & $ -7.7 \pm 0.5$& $0.97 \pm 0.02$ & $147.6 \pm 10.0$ &&          &       &&          &       & Mons \\
54887.567 & $ -7.5 \pm 0.3$& $0.94 \pm 0.01$ & $171.4 \pm 10.0$ &&          &       &&          &       & Mons \\
54887.581 & $ -7.4 \pm 0.4$& $0.90 \pm 0.01$ & $165.6 \pm 10.0$ &&          &       &&          &       & Mons \\
54896.515 & $ -7.3 \pm 0.2$& $0.96 \pm 0.01$ & $161.0 \pm 10.0$ &&          &       &&          &       & Mons \\ 
54896.532 & $ -7.6 \pm 0.3$& $0.97 \pm 0.01$ & $156.8 \pm 10.0$ &&          &       &&          &       & Mons \\
54896.546 & $ -7.4 \pm 0.5$& $0.99 \pm 0.02$ & $158.8 \pm 10.0$ &&          &       &&          &       & Mons \\
54904.587 & $ -7.5 \pm 0.5$& $1.06 \pm 0.02$ & $141.9 \pm 10.0$ &&          &       &&          &       & Mons \\
54846.575 & $ -8.3 \pm 0.2$& $0.98 \pm 0.01$ & $142.9 \pm 4.5$ &&          &       && $ -1.10 \pm 0.19$& $291.8 \pm 5.5$& BeSS \\
55270.392 & $ -6.5 \pm 1.2$& $0.90 \pm 0.03$ & $210.7 \pm 4.5$ && $ -0.14 \pm 0.90$& $253.0 \pm 5.5$&&          &       & BeSS \\
55296.398 & $ -6.9 \pm 0.4$& $0.96 \pm 0.01$ & $169.7 \pm 5.5$ &&          &       &&          &       & BeSS \\
55315.401 & $ -9.4 \pm 1.5$& $0.83 \pm 0.04$ & $201.2 \pm 5.5$ &&          &       &&          &       & BeSS \\
55644.434 & $ -7.0 \pm 0.2$& $0.98 \pm 0.01$ & $177.2 \pm 5.5$ &&          &       &&          &       & BeSS \\
56003.642 & $ -4.6 \pm 1.5$& $0.91 \pm 0.04$ & $204.1 \pm 5.5$ &&          &       &&          &       & BeSS \\
56385.427 & $ -5.0 \pm 1.5$& $1.05 \pm 0.06$ & $222.0 \pm 5.5$ &&          &       &&          &       & BeSS \\
56387.629 & $ -5.6 \pm 0.6$& $0.93 \pm 0.02$ & $227.7 \pm 5.5$ &&          &       &&          &       & BeSS \\
56746.565 & $ -5.4 \pm 0.9$& $0.98 \pm 0.03$ & $217.7 \pm 5.5$ &&          &       &&         &       & BeSS \\
57094.355 & $ -9.6 \pm 1.0$& $0.96 \pm 0.03$ & $203.6 \pm 5.5$ &&          &       &&         &       & BeSS \\
57094.568 & $ -8.0 \pm 0.9$& $1.01 \pm 0.03$ & $201.9 \pm 5.5$ &&          &       &&         &       & BeSS \\
57121.515 & $ -8.0 \pm 0.7$& $0.99 \pm 0.02$ & $192.3 \pm 5.5$ &&          &       &&         &       & BeSS \\
57489.350 & $ -9.3 \pm 0.5$& $0.95 \pm 0.02$ & $173.5 \pm 5.5$ &&          &       &&         &       & BeSS \\
57495.589 & $ -7.1 \pm 0.7$& $0.95 \pm 0.03$ & $169.3 \pm 5.5$ &&          &       &&         &       & BeSS \\
57803.483 & $ -11.7 \pm 0.6$& $1.00 \pm 0.03$ & $162.9 \pm 4.5$ && $ -1.06 \pm 0.50$& $210.7 \pm 5.5$&& $ -1.04 \pm 0.50$& $248.1 \pm 5.5$& BeSS \\
57840.371 & $ -11.9 \pm 0.9$& $1.01 \pm 0.01$ & $177.4 \pm 5.5$ &&          &       &&         &       & BeSS \\
58060.678 & $ -12.3 \pm 0.5$& $0.97 \pm 0.03$ & $166.5 \pm 4.5$ && $ -1.27 \pm 0.50$& $216.8 \pm 5.5$&& $ -1.16 \pm 0.50$& $255.9 \pm 5.5$& BeSS \\
\hline
  \end{tabular}
\end{table*}
\addtocounter{table}{-1}
\begin{table*}
  \caption{Continued}
  \small
  \begin{tabular}{c r c c c c c c c c c}
    \hline
HJD-2\,400\,000 &  \multicolumn{3}{c}{H$\alpha$} & & \multicolumn{2}{c}{H$\beta$} & &\multicolumn{2}{c}{He\,{\sc i} $\lambda$\,5876} & Instrument \\
\cline{2-4}\cline{6-7}\cline{9-10}
                &  EW & V/R & $\Delta\,v$ & & EW & $\Delta\,v$ & & EW & $\Delta\,v$ \\ 
& (\AA) & & (km\,s$^{-1}$) & & (\AA) & (km\,s$^{-1}$) & & (\AA) & (km\,s$^{-1}$) \\
\hline
58180.437 & $ -13.3 \pm 0.7$& $0.96 \pm 0.02$ & $154.0 \pm 4.5$ && $ -1.61 \pm 0.55$& $195.0 \pm 5.5$ && $ -1.56 \pm 0.55$& $241.2 \pm 5.5$& BeSS \\
58199.598 & $ -13.3 \pm 1.0$& $0.94 \pm 0.03$ & $171.5 \pm 5.5$ &&          &       &&         &       & BeSS \\
58541.294 & $ -16.5 \pm 1.6$& $0.96 \pm 0.03$ & $177.2 \pm 5.5$ &&          &       &&         &       & BeSS \\
58565.437 & $ -7.9 \pm 1.5$& $1.03 \pm 0.03$ & $175.7 \pm 5.5$ &&          &       &&         &       & BeSS \\
58594.614 & $ -7.0 \pm 0.9$& $1.08 \pm 0.03$ & $186.4 \pm 5.5$ &&          &       &&         &       & BeSS \\
55573.218 & $ -5.6 \pm 0.2$& $1.03 \pm 0.01$ & $217.3 \pm 1.5$ && $ -0.06 \pm 0.16$& $265.6 \pm 2.9$&& $ -0.36 \pm 0.21$& $316.0 \pm 2.5$ & NOT \\
55577.227 & $ -5.5 \pm 0.3$& $0.95 \pm 0.01$ & $217.1 \pm 1.5$ && $ +0.00 \pm 0.16$& $270.0 \pm 2.9$ && $ -0.36 \pm 0.21$& $318.0 \pm 2.5$ & NOT \\
56285.721 & $ -5.7 \pm 0.2$& $1.05 \pm 0.01$ & $214.8 \pm 1.5$ && $ +0.47 \pm 0.13$& $276.7 \pm 2.9$&& $ -0.47 \pm 0.21$& $326.6 \pm 2.5$ & NOT \\
56019.496 & $ -4.5 \pm 0.4$& $0.93 \pm 0.02$ & $207.0 \pm 1.5$ && $ +0.64 \pm 0.29$& $301.8 \pm 3.0$&& $ -0.43 \pm 0.35$& $401.3 \pm 2.5$ & Coralie \\
56603.958 & $ -6.2 \pm 0.2$& $1.00 \pm 0.01$ & $211.3 \pm 3.0$ && $ +0.12 \pm 0.24$& $267.5 \pm 2.9$ && $ -0.75 \pm 0.11$& $314.6 \pm 2.9$ & TIGRE \\
56641.891 & $ -6.4 \pm 0.2$& $1.00 \pm 0.01$ & $218.8 \pm 3.0$ && $ +0.19 \pm 0.15$& $267.5 \pm 2.9$ && $ -0.62 \pm 0.10$& $309.9 \pm 2.9$ & TIGRE \\
56642.851 & $ -6.4 \pm 0.2$& $1.03 \pm 0.01$ & $218.4 \pm 3.0$ && $ +0.15 \pm 0.14$& $265.2 \pm 2.9$ && $ -0.68 \pm 0.08$& $321.6 \pm 2.9$ & TIGRE \\
56643.860 & $ -6.1 \pm 0.2$& $1.02 \pm 0.01$ & $209.5 \pm 3.0$ && $ +0.44 \pm 0.16$& $263.3 \pm 2.9$ && $ -0.65 \pm 0.08$& $299.1 \pm 2.9$ & TIGRE \\
56644.860 & $ -6.3 \pm 0.1$& $1.01 \pm 0.01$ & $214.0 \pm 3.0$ && $ +0.13 \pm 0.17$& $259.9 \pm 2.9$ && $ -0.68 \pm 0.08$& $307.3 \pm 2.9$ & TIGRE \\
56674.773 & $ -5.6 \pm 0.1$& $1.02 \pm 0.01$ & $212.1 \pm 3.0$ && $ +0.44 \pm 0.15$& $272.2 \pm 2.9$ && $ -0.52 \pm 0.07$& $313.3 \pm 2.9$ & TIGRE \\
56679.826 & $ -5.9 \pm 0.1$& $0.94 \pm 0.01$ & $212.1 \pm 3.0$ && $ -0.03 \pm 0.23$& $263.2 \pm 2.9$ && $ -0.29 \pm 0.08$& $312.4 \pm 2.9$ & TIGRE \\
56680.791 & $ -5.7 \pm 0.2$& $0.95 \pm 0.01$ & $212.6 \pm 3.0$ && $ +0.15 \pm 0.15$& $262.8 \pm 2.9$ && $ -0.52 \pm 0.06$& $319.3 \pm 2.9$ & TIGRE \\
56713.657 &                &                 &                 && $ +0.80 \pm 0.31$& $281.6 \pm 2.9$ && $ -0.59 \pm 0.11$& $321.2 \pm 2.9$ & TIGRE \\
56948.991 & $ -6.3 \pm 0.3$& $0.93 \pm 0.01$ & $215.3 \pm 3.0$ && $ -0.16 \pm 0.30$& $252.0 \pm 2.9$ && $ -0.36 \pm 0.11$& $298.7 \pm 2.9$ & TIGRE \\
57006.869 & $ -5.7 \pm 0.4$& $0.93 \pm 0.02$ & $195.7 \pm 3.0$ && $ -0.13 \pm 0.33$& $243.3 \pm 2.9$ && $ -0.08 \pm 0.30$& $302.7 \pm 2.9$ & TIGRE \\
57038.773 & $ -8.2 \pm 0.4$& $1.01 \pm 0.01$ & $195.0 \pm 3.0$ && $ -0.37 \pm 0.46$& $237.6 \pm 2.9$ && $ -0.70 \pm 0.27$& $273.3 \pm 2.9$ & TIGRE \\
57073.668 & $ -7.2 \pm 0.3$& $1.02 \pm 0.01$ & $200.4 \pm 3.0$ && $ -0.54 \pm 0.22$& $243.0 \pm 2.9$ &&          &       & TIGRE \\
57105.637 & $ -7.3 \pm 0.5$& $1.00 \pm 0.02$ & $189.9 \pm 3.0$ && $ -0.35 \pm 0.28$& $229.5 \pm 2.9$ && $ -0.51 \pm 0.25$& $272.7 \pm 2.9$ & TIGRE \\
57316.991 & $ -8.0 \pm 0.1$& $1.00 \pm 0.01$ & $189.6 \pm 3.0$ && $ -0.83 \pm 0.22$& $229.7 \pm 2.9$ && $ -0.62 \pm 0.08$& $257.4 \pm 2.9$ & TIGRE \\
57374.853 & $ -8.3 \pm 0.6$& $1.01 \pm 0.02$ & $179.1 \pm 3.0$ &&          &       &&          &       & TIGRE \\
57435.664 & $ -7.7 \pm 0.5$& $0.95 \pm 0.02$ & $189.9 \pm 3.0$ &&          &       &&          &       & TIGRE \\
57439.657 & $ -7.7 \pm 0.2$& $0.97 \pm 0.01$ & $187.0 \pm 3.0$ && $ -0.39 \pm 0.29$& $213.7 \pm 2.9$ && $ -0.28 \pm 0.10$& $262.9 \pm 2.9$ & TIGRE \\
57799.692 & $ -12.0 \pm 0.2$& $1.05 \pm 0.01$ & $159.1 \pm 3.0$ && $ -1.25 \pm 0.14$& $202.6 \pm 2.9$ && $ -1.05 \pm 0.13$& $246.2 \pm 2.9$ & TIGRE \\
57849.599 & $ -11.4 \pm 0.3$& $1.02 \pm 0.01$ & $172.5 \pm 3.0$ && $ -1.18 \pm 0.39$& $207.5 \pm 2.9$ && $ -0.95 \pm 0.14$& $229.9 \pm 2.9$ & TIGRE \\
58067.949 & $ -12.4 \pm 0.3$& $0.93 \pm 0.01$ & $173.6 \pm 3.0$ && $ -1.03 \pm 0.21$& $214.4 \pm 2.9$ && $ -1.34 \pm 0.09$& $245.2 \pm 2.9$ & TIGRE \\
58070.887 & $ -12.2 \pm 0.3$& $0.97 \pm 0.01$ & $174.3 \pm 3.0$ && $ -0.96 \pm 0.36$& $215.4 \pm 2.9$ && $ -1.34 \pm 0.16$& $240.8 \pm 2.9$ & TIGRE \\
58072.978 & $ -12.3 \pm 0.2$& $0.97 \pm 0.01$ & $170.3 \pm 3.0$ && $ -1.07 \pm 0.26$& $209.2 \pm 2.9$ && $ -1.25 \pm 0.09$& $234.9 \pm 2.9$ & TIGRE \\
58074.879 & $ -12.3 \pm 0.2$& $0.98 \pm 0.01$ & $171.1 \pm 3.0$ && $ -1.04 \pm 0.21$& $211.5 \pm 2.9$ && $ -1.21 \pm 0.11$& $240.7 \pm 2.9$ & TIGRE \\
58151.729 & $ -12.8 \pm 0.3$& $0.95 \pm 0.01$ & $167.5 \pm 3.0$ && $ -1.31 \pm 0.34$& $206.9 \pm 2.9$ && $ -1.13 \pm 0.10$& $238.3 \pm 2.9$ & TIGRE \\
58188.698 & $ -13.3 \pm 0.2$& $0.97 \pm 0.01$ & $164.3 \pm 3.0$ && $ -1.40 \pm 0.24$& $203.3 \pm 2.9$ && $ -1.51 \pm 0.12$& $251.4 \pm 2.9$ & TIGRE \\
58218.604 & $ -13.6 \pm 0.3$& $1.05 \pm 0.01$ & $166.8 \pm 3.0$ && $ -1.27 \pm 0.27$& $198.9 \pm 2.9$ && $ -1.61 \pm 0.11$& $229.4 \pm 2.9$ & TIGRE \\
58389.980 & $ -13.0 \pm 0.5$& $1.02 \pm 0.02$ & $170.7 \pm 3.0$ && $ -1.09 \pm 0.39$& $208.9 \pm 2.9$ && $ -2.08 \pm 0.15$& $236.5 \pm 2.9$ & TIGRE \\
58446.863 & $ -12.7 \pm 0.3$& $0.94 \pm 0.01$ & $172.2 \pm 3.0$ && $ -1.11 \pm 0.27$& $211.0 \pm 2.9$ && $ -1.67 \pm 0.14$& $238.8 \pm 2.9$ & TIGRE \\
58466.813 & $ -11.5 \pm 0.2$& $1.03 \pm 0.01$ & $175.0 \pm 3.0$ && $ -1.06 \pm 0.35$& $204.5 \pm 2.9$ && $ -1.25 \pm 0.08$& $229.1 \pm 2.9$ & TIGRE \\
58471.786 & $ -12.1 \pm 0.3$& $1.09 \pm 0.01$ & $169.5 \pm 3.0$ && $ -0.95 \pm 0.95$& $206.2 \pm 2.9$ && $ -1.21 \pm 0.27$& $235.5 \pm 2.9$ & TIGRE \\
58477.813 & $ -12.0 \pm 0.4$& $1.03 \pm 0.01$ & $170.3 \pm 3.0$ && $ -1.26 \pm 0.60$& $195.0 \pm 2.9$ && $ -1.31 \pm 0.39$& $229.8 \pm 2.9$ & TIGRE \\
58478.814 & $ -11.7 \pm 0.4$& $1.01 \pm 0.01$ & $169.5 \pm 3.0$ && $ -1.14 \pm 0.47$& $194.8 \pm 2.9$ && $ -1.06 \pm 0.16$& $224.3 \pm 2.9$ & TIGRE \\
58481.767 & $ -11.4 \pm 0.5$& $1.00 \pm 0.01$ & $171.7 \pm 3.0$ && $ -0.96 \pm 0.83$& $204.4 \pm 2.9$ && $ -0.95 \pm 0.21$& $232.1 \pm 2.9$ & TIGRE \\
58490.801 & $ -11.5 \pm 0.2$& $1.03 \pm 0.01$ & $164.3 \pm 3.0$ && $ -1.24 \pm 0.56$& $197.5 \pm 2.9$ && $ -1.11 \pm 0.15$& $221.9 \pm 2.9$ & TIGRE \\
58509.730 & $ -10.7 \pm 0.2$& $1.00 \pm 0.01$ & $169.5 \pm 3.0$ && $ -0.79 \pm 0.37$& $202.5 \pm 2.9$ && $ -0.94 \pm 0.09$& $240.8 \pm 2.9$ & TIGRE \\
58526.766 & $ -10.7 \pm 0.2$& $0.93 \pm 0.01$ & $172.6 \pm 3.0$ && $ -0.86 \pm 0.37$& $206.0 \pm 2.9$ && $ -1.06 \pm 0.08$& $244.5 \pm 2.9$ & TIGRE \\
58556.678 & $ -9.4 \pm 0.2$& $0.96 \pm 0.01$ & $168.5 \pm 3.0$ && $ -0.45 \pm 0.31$& $212.7 \pm 2.9$ && $ -0.89 \pm 0.12$& $254.5 \pm 2.9$ & TIGRE \\
58759.982 & $ -7.0 \pm 0.2$& $0.80 \pm 0.01$ & $196.5 \pm 3.0$ && $ -0.20 \pm 0.20$& $269.4 \pm 2.9$ && $ -0.78 \pm 0.12$& $325.9 \pm 2.9$ & TIGRE \\
58790.962 & $ -7.3 \pm 0.2$& $1.06 \pm 0.01$ & $194.7 \pm 3.0$ && $ +0.14 \pm 0.21$& $246.9 \pm 2.9$ && $ -0.65 \pm 0.14$& $315.9 \pm 2.9$ & TIGRE \\
58820.807 & $ -5.6 \pm 0.2$& $1.09 \pm 0.01$ & $199.6 \pm 3.0$ && $ +0.47 \pm 0.19$& $271.4 \pm 2.9$ && $ -0.30 \pm 0.10$& $355.0 \pm 2.9$ & TIGRE \\
58852.795 & $ -5.0 \pm 0.2$& $0.98 \pm 0.01$ & $193.2 \pm 3.0$ && $ +0.63 \pm 0.23$& $303.6 \pm 2.9$ && $ -0.30 \pm 0.12$& $386.1 \pm 2.9$ & TIGRE \\
58877.868 & $ -4.2 \pm 0.2$& $0.93 \pm 0.01$ & $200.9 \pm 3.0$ && $ +0.90 \pm 0.21$& $330.4 \pm 2.9$ && $ -0.16 \pm 0.09$& $415.5 \pm 2.9$ & TIGRE \\
58900.628 & $ -4.1 \pm 0.2$& $0.94 \pm 0.01$ & $218.9 \pm 3.0$ && $ +0.62 \pm 0.28$& $301.4 \pm 2.9$ && $ -0.22 \pm 0.12$& $405.8 \pm 2.9$ & TIGRE \\
58446.542 & $ -13.3 \pm 0.5$& $0.99 \pm 0.02$ & $161.9 \pm 10.0$ &&          &       &&          &       & FC \\
58461.521 & $ -11.2 \pm 0.6$& $1.02 \pm 0.02$ & $154.3 \pm 10.0$ &&          &       &&          &       & FC \\
58469.422 & $ -11.6 \pm 0.7$& $1.12 \pm 0.02$ & $135.7 \pm 10.0$ &&          &       &&          &       & FC \\
58481.470 & $ -12.0 \pm 0.5$& $1.02 \pm 0.02$ & $146.2 \pm 10.0$ &&          &       &&          &       & FC \\
\hline
  \end{tabular}
\end{table*}
\addtocounter{table}{-1}
\begin{table*}
  \caption{Continued}
  \small
  \begin{tabular}{c r c c c c c c c c c}
    \hline
HJD-2\,400\,000 &  \multicolumn{3}{c}{H$\alpha$} & & \multicolumn{2}{c}{H$\beta$} & &\multicolumn{2}{c}{He\,{\sc i} $\lambda$\,5876} & Instrument \\
\cline{2-4}\cline{6-7}\cline{9-10}
                &  EW & V/R & $\Delta\,v$ & & EW & $\Delta\,v$ & & EW & $\Delta\,v$ \\ 
& (\AA) & & (km\,s$^{-1}$) & & (\AA) & (km\,s$^{-1}$) & & (\AA) & (km\,s$^{-1}$) \\
\hline

58494.416 & $ -11.3 \pm 0.3$& $1.01 \pm 0.01$ & $143.1 \pm 10.0$ &&          &       &&          &       & FC \\
58525.423 & $ -10.5 \pm 0.7$& $0.96 \pm 0.02$ & $157.4 \pm 10.0$ &&          &       &&          &       & FC \\
58571.369 & $ -8.4 \pm 0.3$& $0.99 \pm 0.01$ & $162.0 \pm 10.0$ &&          &       &&          &       & FC \\
58606.356 & $ -7.2 \pm 1.0$& $1.03 \pm 0.03$ & $150.4 \pm 10.0$ &&          &       &&          &       & FC \\
58456.499 & $ -10.8 \pm 0.5 $& $1.01 \pm 0.02$ & $173.0 \pm 4.5$ &&          &       && $ -1.35 \pm 0.38$& $236.3 \pm 3.5$ & JGF \\
58473.503 & $ -10.6 \pm 0.4$& $1.09 \pm 0.01$ & $170.1 \pm 4.5$ &&          &       && $ -1.39 \pm 0.23$& $227.4 \pm 3.5$ & JGF \\
58491.487 & $ -10.4 \pm 0.4$& $1.03 \pm 0.01$ & $168.3 \pm 4.5$ &&          &       && $ -0.93 \pm 0.22$& $230.9 \pm 3.5$ & JGF \\
58495.507 & $ -10.7 \pm 0.4$& $0.98 \pm 0.01$ & $168.8 \pm 4.5$ &&          &       && $ -1.12 \pm 0.23$& $231.7 \pm 3.5$ & JGF \\
58509.475 & $ -10.0 \pm 0.4$& $1.00 \pm 0.01$ & $169.7 \pm 4.5$ &&          &       &&          &       & JGF \\
58520.456 & $ -9.5 \pm 0.3$& $0.97 \pm 0.01$ & $170.8 \pm 4.5$ &&          &       && $ -1.02 \pm 0.22$& $236.3 \pm 3.5$ & JGF \\
58522.477 & $ -9.8 \pm 0.4$& $0.98 \pm 0.01$ & $171.5 \pm 4.5$ &&          &       &&          &       & JGF \\
58528.455 & $ -9.6 \pm 0.3$& $0.96 \pm 0.01$ & $171.5 \pm 4.5$ &&          &       && $ -1.01 \pm 0.17$& $245.8 \pm 3.5$ & JGF \\
58535.449 & $ -9.1 \pm 0.3$& $1.03 \pm 0.01$ & $172.8 \pm 4.5$ &&          &       && $ -0.90 \pm 0.20$& $239.6 \pm 3.5$ & JGF \\
58541.474 & $ -9.4 \pm 0.5$& $0.94 \pm 0.02$ & $174.3 \pm 4.5$ &&          &       && $ -1.01 \pm 0.30$& $254.4 \pm 3.5$ & JGF \\
58550.468 & $ -8.4 \pm 0.4$& $0.91 \pm 0.01$ & $173.8 \pm 4.5$ &&          &       &&          &       & JGF \\
58555.422 & $ -8.1 \pm 0.5$& $0.91 \pm 0.02$ & $172.0 \pm 4.5$ &&          &       &&          &       & JGF \\
58571.429 & $ -8.0 \pm 0.3$& $0.97 \pm 0.01$ & $173.4 \pm 4.5$ &&          &       && $ -0.58 \pm 0.21$& $241.1 \pm 3.5$ & JGF \\
58578.339 & $ -8.2 \pm 0.3$& $0.96 \pm 0.01$ & $177.3 \pm 4.5$ &&          &       && $ -0.59 \pm 0.21$& $292.5 \pm 3.5$ & JGF \\
58485.421 & $ -11.0 \pm 0.5$& $0.99 \pm 0.02$ & $166.9 \pm 4.5$ &&          &       &&         &       & US \\
58490.357 & $ -10.8 \pm 0.5$& $1.01 \pm 0.02$ & $168.3 \pm 4.5$ &&          &       &&         &       & US \\
\hline
  \end{tabular}
\end{table*}

\subsection{Archival photometric data}
The database of the American Association of Variable Star Observers\footnote{https://www.aavso.org} \citep[AAVSO,][]{Kaf16} contains over 4400 visual magnitude measurements of HD~60\,848 obtained by many different amateur observers between 1939 and 2018 with the bulk of the data taken between 1950 and 2002.

We also retrieved $V$-band photometry of HD~60\,848 taken from the All Sky Automated Survey \citep[ASAS-3,][]{ASAS}\footnote{http://www.astrouw.edu.pl/$\sim$gp/asas/} and from the All-Sky Automated Survey for Supernovae \citep[ASAS-SN][]{ASAS-SN}\footnote{https://asas-sn.osu.edu/}. The ASAS-3 data were collected at the Las Campanas Observatory in Chile using two wide-field ($8.8^{\circ} \times 8.8^{\circ}$) telescopes, each equipped with a 200/2.8 Minolta telephoto lens and a $2048 \times 2048$ pixels AP-10 CCD camera. These instruments were complemented by a 25\,cm Cassegrain telescope equipped with the same type of CCD camera covering a narrower field ($2.2^{\circ} \times 2.2^{\circ}$). ASAS-3 provides aperture photometry for five different apertures varying in diameter from 2 to 6 pixels. 
The ASAS-3 data are not uniform in terms of exposure time and thus in terms of saturation limit, as the exposure time was changed in the course of the project from 180\,s (saturation near $V \sim 7.5$) to 60\,s (saturation around $V \sim 6$). Given the magnitude of HD~60\,848 ($V = 6.87$ in SIMBAD), we filtered the observations, keeping only the best data (grade A). We further applied a median absolute deviation (MAD) filtering which consists in discarding values that deviate by more than $3 \times {\rm median}(|V_i - {\rm median}(V_i)|)$ where $V_i$ is the ASAS-3 $V$-band magnitude. The ASAS-SN data span the years 2012 -- 2018, i.e.\ they cover the most recent cycle of HD~60\,848, except for the recent steep decay phase. The saturation limit of the ASAS-SN data is around magnitude 10 -- 11 depending on the camera used for the observation. However, for relatively isolated sources (which is the case for HD~60\,848), photons lost due to saturation are recovered in the data processing \citep{ASAS-SN}. Yet, this reconstruction is not perfect and for an object as bright as HD~60\,848, the reconstructed photometry must be considered with caution \citep{ASAS-SN}. In our case, the largest dataset (800 measurements) is provided by the {\tt bb} camera at the Haleakala Observatory (Hawaii). We thus focused our analysis on the data from this camera. To remove outliers, we applied the same MAD filtering as for the ASAS-3 data.

Finally, we have also retrieved the $H_p$ photometry of HD~60\,848 collected by ESA's {\it Hipparcos} satellite between March 1990 and March 1993 \citep{ESA97}.

\subsection{X-ray observations}        
Since our optical monitoring revealed that the star had reached a plateau in the emission line strengths by the end of 2018 (see Sect.\,\ref{results} hereafter), we triggered a Target of Opportunity observation with the {\it Chandra} X-ray Observatory. The 18.8\,ks observation was taken on 30 December 2018 using the ACIS-S instrument (ObsID 22012). The level=2 event file provided by the {\it Chandra} pipeline was further processed using CIAO v4.9 and CALDB v4.7.3. The source counts were extracted from a circle of 2.5\,arcsec radius while the background was estimated from a surrounding annulus with radii 2.5 and 12.3\,arcsec. To this aim, we used the CIAO task {\sc specextract} generating first unweighted response matrices but still performing an aperture correction for the ancillary response file (correctpsf=yes), as it is considered appropriate for point source analysis. In a second time, we generated weighted response matrices as a check. As the fit results obtained with both methods were indistinguishable, we only present those obtained with the first approach in this paper. The source spectrum was binned to reach a minimum of 10 counts per bin.
The {\it Chandra} observation was taken to check for possible variations with respect to a previous {\it XMM-Newton} observation taken on 2 April 2012 when the star was in a low H$\alpha$ emission state (EW(H$\alpha$) = $-4.5$\,\AA). Details on the {\it XMM-Newton} observation can be found in \citet{Rau13}.

\section{Results} \label{results}
\subsection{Spectroscopic variability} \label{specvar}
\begin{figure}
 \resizebox{8.5cm}{!}{\includegraphics{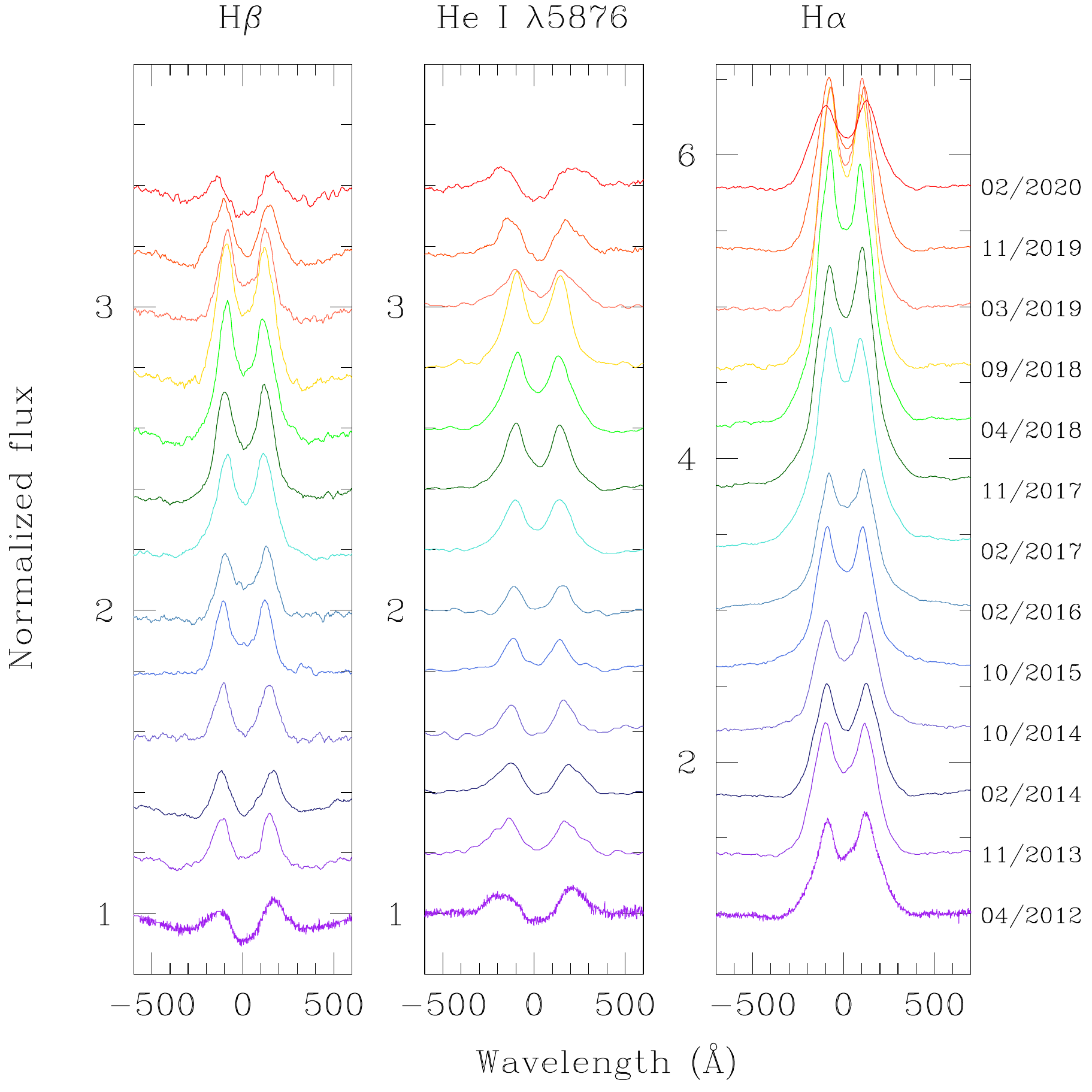}}
\caption{Evolution of the H$\beta$ (left), He\,{\sc i} $\lambda$\,5876 (middle) and H$\alpha$ (right) emission lines as a function of time. The epochs of the observations are indicated on the right. For clarity, the consecutive continuum-normalized spectra are shifted vertically by 0.1 continuum units in the left and middle panel, and by 0.2 continuum units in the rightmost panel. The different lines are shown in heliocentric radial velocity space.}
\label{montage}
\end{figure}

The most prominent emission lines in the optical spectrum of HD~60\,848 are H$\alpha$, H$\beta$ and He\,{\sc i} $\lambda$\,5876. Figure\,\ref{montage} displays the line profiles observed between April 2012 and February 2020. In April 2012, the emission lines were at their minimum strength. We clearly see that the outer wings of the H$\beta$ line, and to a lesser extent also of the H$\alpha$ line were in absorption. As time went by, the first variations concerned these wings which got progressively filled-in by emission. From October 2014 on, the wings were completely in emission. Over subsequent years the emission wings extended to higher velocities (300\,km\,s$^{-1}$ for H$\beta$ and 500\,km\,s$^{-1}$ for H$\alpha$). These extended emission wings most probably reflect an increase in the disk density and thus enhanced scattering by free electrons \citep{Poe79,Lab18}. Between February 2016 and February 2017, the overall strength of the emission tremendously increased. In 2018 (April for H$\beta$ and September for H$\alpha$), the strength of the emission wings started receding again, and in March 2019 the decrease of the emission strength affected the central part of the line profiles. The most recent data from the 2019 -- 2020 observing season reveal a spectrum very similar to what was observed in 2012.  

\begin{figure*}
  \begin{minipage}{5.8cm}
    \begin{center}
      \resizebox{5.8cm}{!}{\includegraphics{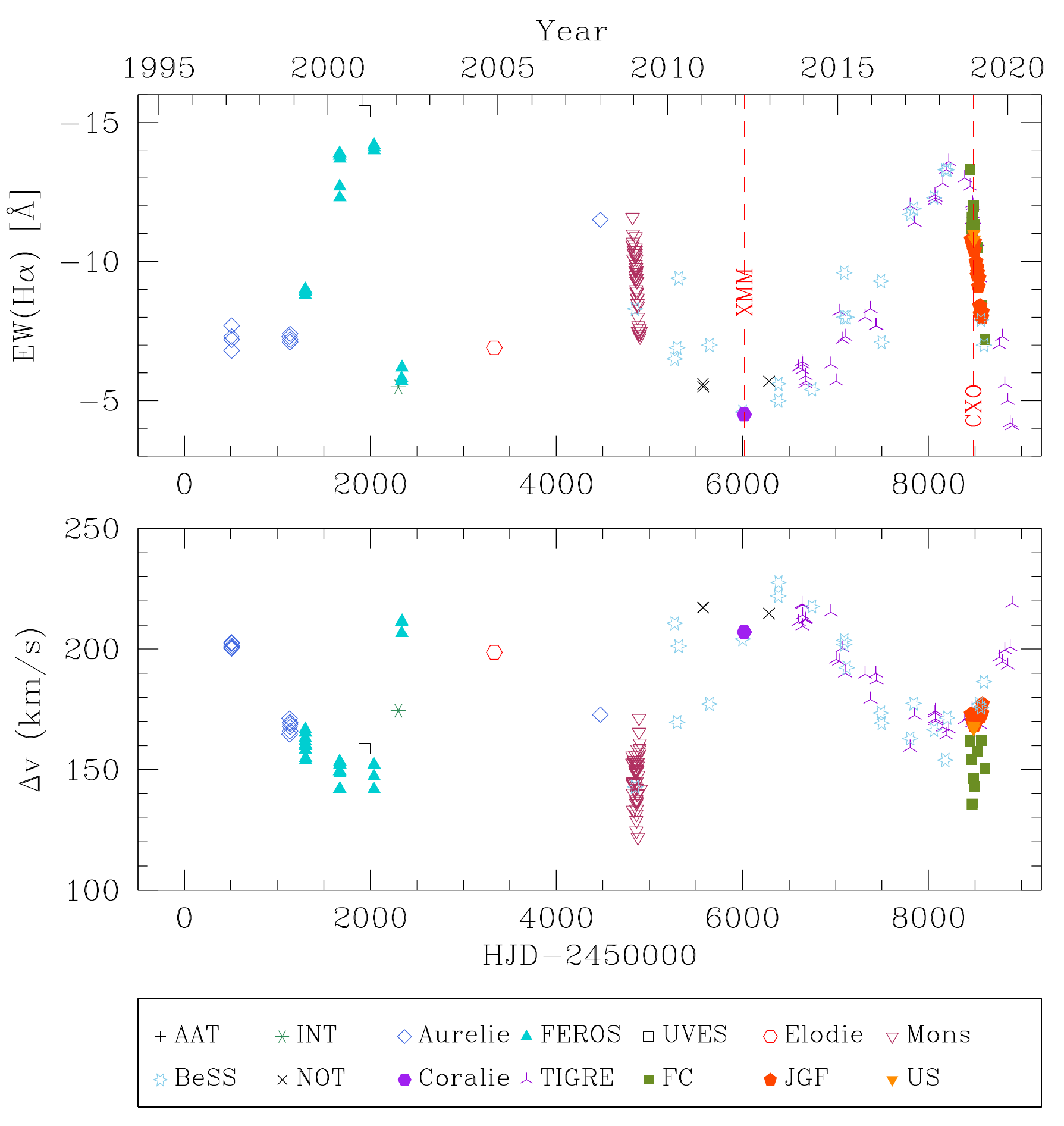}}
    \end{center}
  \end{minipage}
    \begin{minipage}{5.8cm}
    \begin{center}
      \resizebox{5.8cm}{!}{\includegraphics{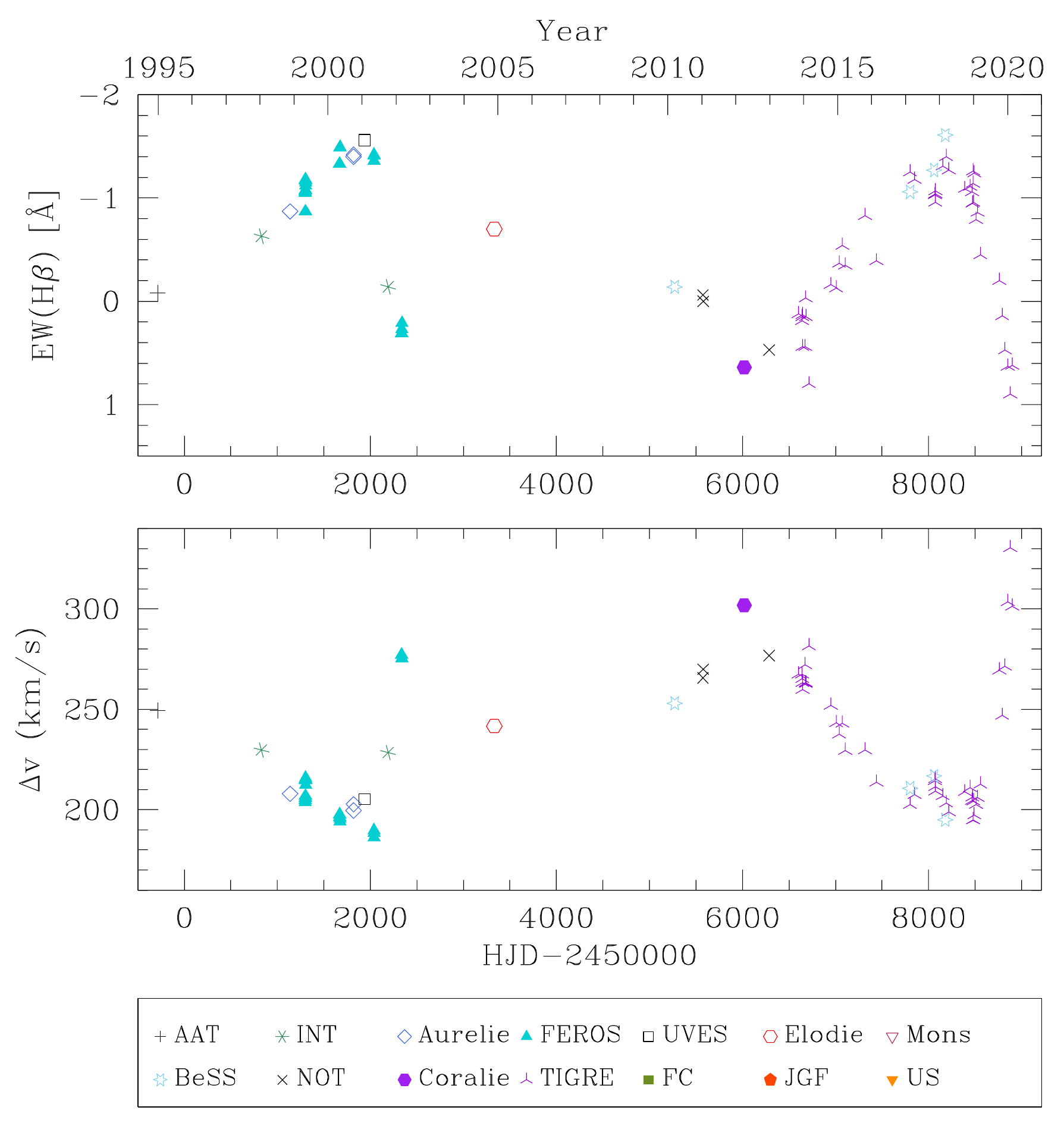}}
    \end{center}
    \end{minipage}
    \begin{minipage}{5.8cm}
    \begin{center}
      \resizebox{5.8cm}{!}{\includegraphics{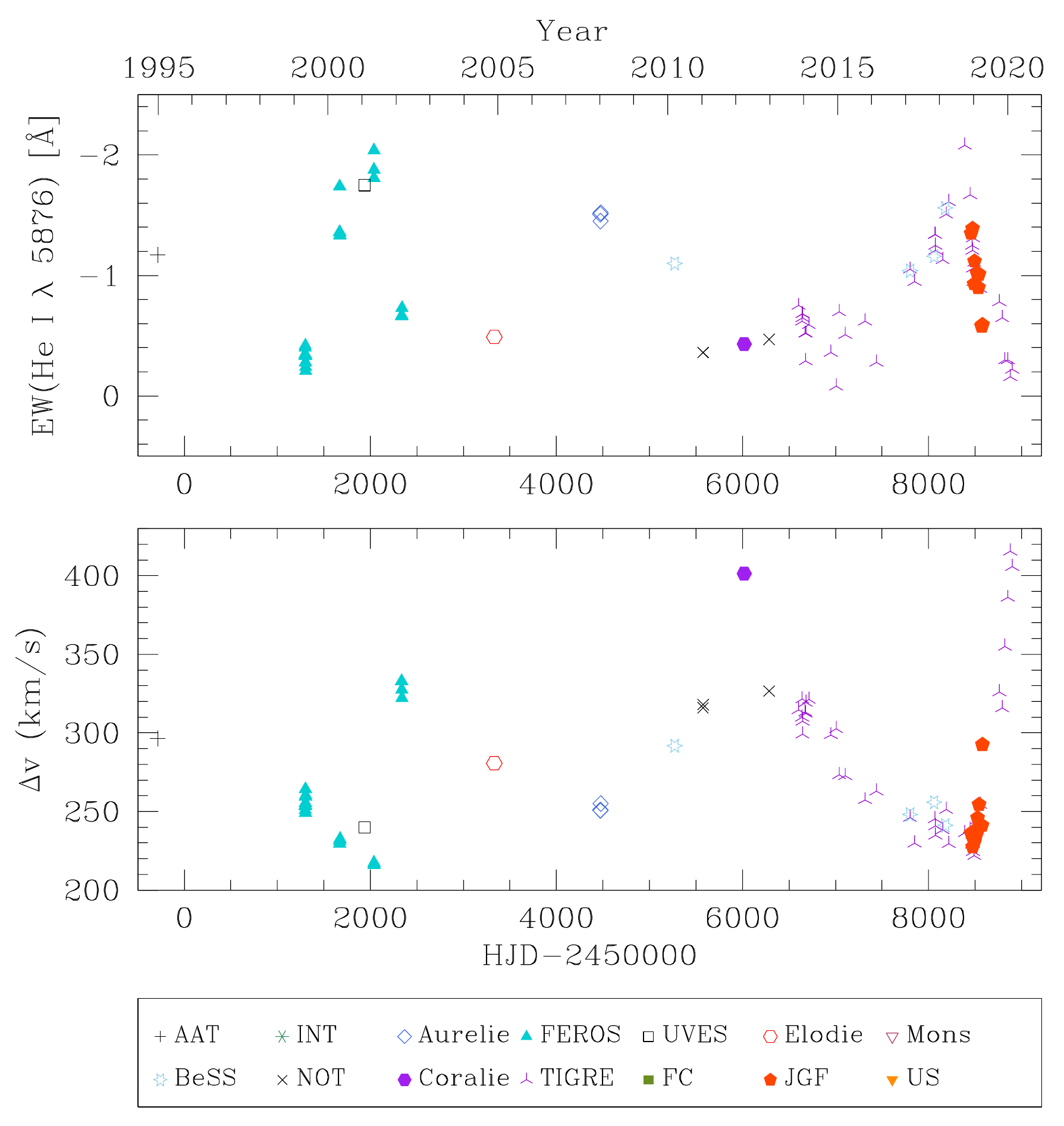}}
    \end{center}
  \end{minipage}
\caption{Evolution of the EWs and $\Delta v$ of the double-peaked H$\alpha$ (left), H$\beta$ (middle) and He\,{\sc i} $\lambda$\,5876 emission lines with time. The different symbols indicate different instruments. The dashed vertical lines in the EW(H$\alpha$) panel indicate the dates of the {\it XMM-Newton} and {\it Chandra} observations.}
\label{EWs}
\end{figure*}

These variations were associated with important variations of the EWs.
For instance, the EW of H$\alpha$ varied between about $-4$\,\AA\ at minimum emission state and $-15$\,\AA\ at maximum emission (see Fig.\,\ref{EWs}). The variations of the EWs of the H$\beta$ and He\,{\sc i} $\lambda$\,5876 lines are qualitatively similar, though the H$\beta$ line has a positive (i.e.\ absorption-dominated) EW at minimum emission state. In this line, the disk emission at minimum lies on top of a very broad photospheric absorption. 

All prominent emission lines display double-peaked morphologies on all our observations. Although both emission peaks are always of comparable strengths, the relative intensity of the violet and red emission peaks as well as their separation slightly change with time. Figure\,\ref{EWs} illustrates the variations of the velocity separation of the peaks as a function of time. For the H$\alpha$ line, the largest separation (about 220\,km\,s$^{-1}$) is observed when the emission is low. This velocity separation reduces to about 150\,km\,s$^{-1}$ at phases of maximum emission.

\begin{figure}
    \begin{center}
      \resizebox{9cm}{!}{\includegraphics{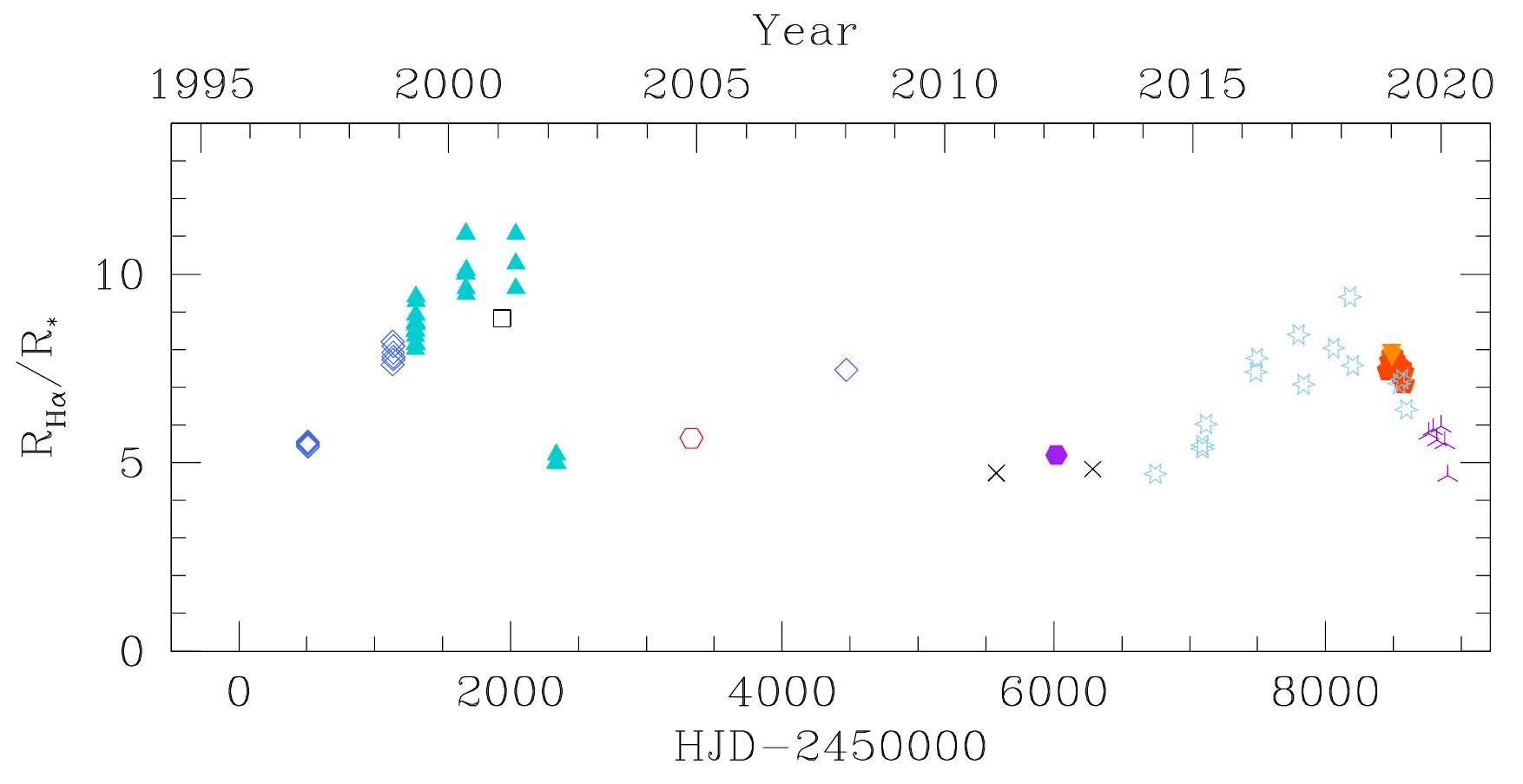}}
    \end{center}
\caption{Evolution of the outer radius of the H$\alpha$ emission region computed from the $\Delta\,v$ values using the formalism of \citet{Zam19}. The symbols have the same meaning as in Fig.\,\ref{EWvsdv}. To avoid biases, only observations with a resolving power of 8000 or higher are shown.}
\label{radiusHa}
\end{figure}

This behaviour is expected for emissions arising from a Keplerian circumstellar disk \citep[e.g.][]{HV95}. Indeed, for an optically thin  Keplerian disk located in the stellar equatorial plane, the velocity separation of the peaks of nearly symmetrical H$\alpha$ emission lines reflects to first order the radius of the line emitting region \citep[e.g.][]{HV95,Cat13,Zam19}:
\begin{equation}
  R_{H\alpha} \propto R_*\,\left(\frac{2\,v\,\sin{i}}{\Delta\,v}\right)^2
  \label{eq1}
\end{equation}
At first sight, the observed variations of $\Delta\,v$ over the cycles of HD~60\,848 could thus simply reflect an increase of the characteristic radius of the H$\alpha$ emitting region by a factor $\sim 2.2$ as the disk goes from minimum emission state to maximum (see Fig.\,\ref{radiusHa}). Alternatively, the separation between the peaks could also vary in response to changes of the apparent inclination of the disk structure if the disk is tilted or warped \citep{Hum98,Por98}. In the case of HD~60\,848, this explanation is however unlikely as we see no triple peaks or other transient features in the line profiles that could hint at a tilted or warped disk. 

\begin{figure*}
  \begin{minipage}{5.8cm}
    \begin{center}
      \resizebox{5.8cm}{!}{\includegraphics{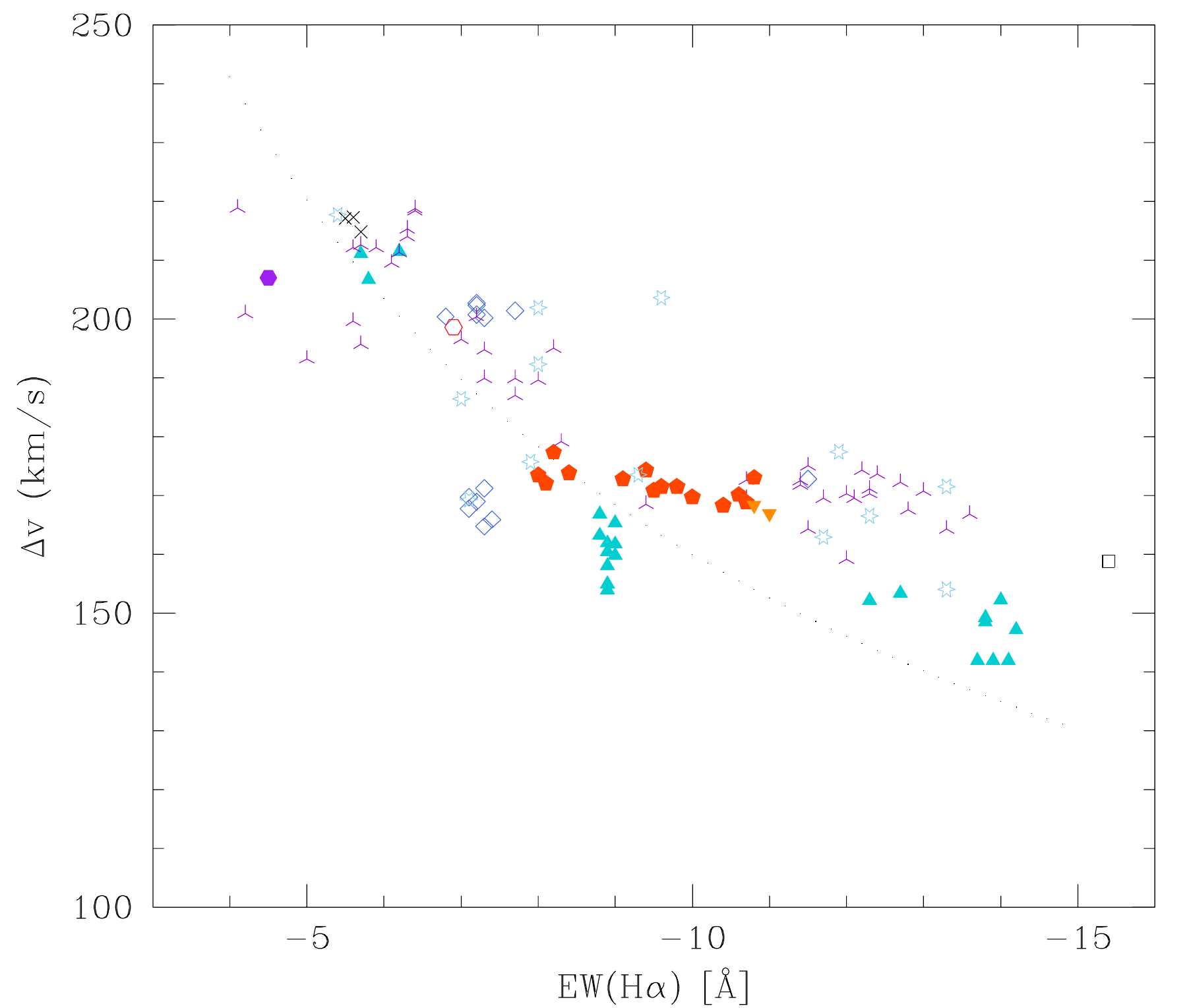}}
    \end{center}
  \end{minipage}
    \begin{minipage}{5.8cm}
    \begin{center}
      \resizebox{5.8cm}{!}{\includegraphics{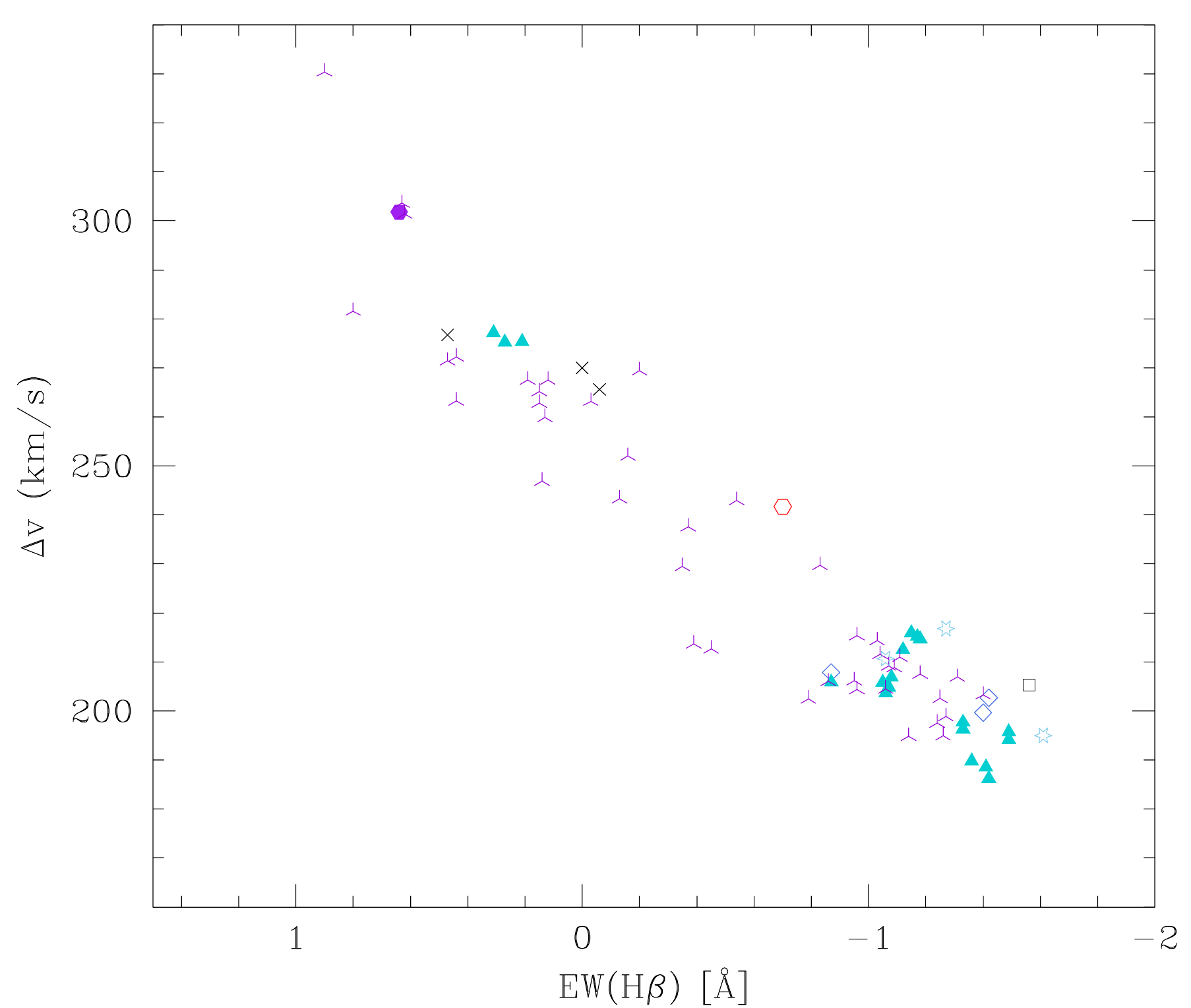}}
    \end{center}
    \end{minipage}
    \begin{minipage}{5.8cm}
    \begin{center}
      \resizebox{5.8cm}{!}{\includegraphics{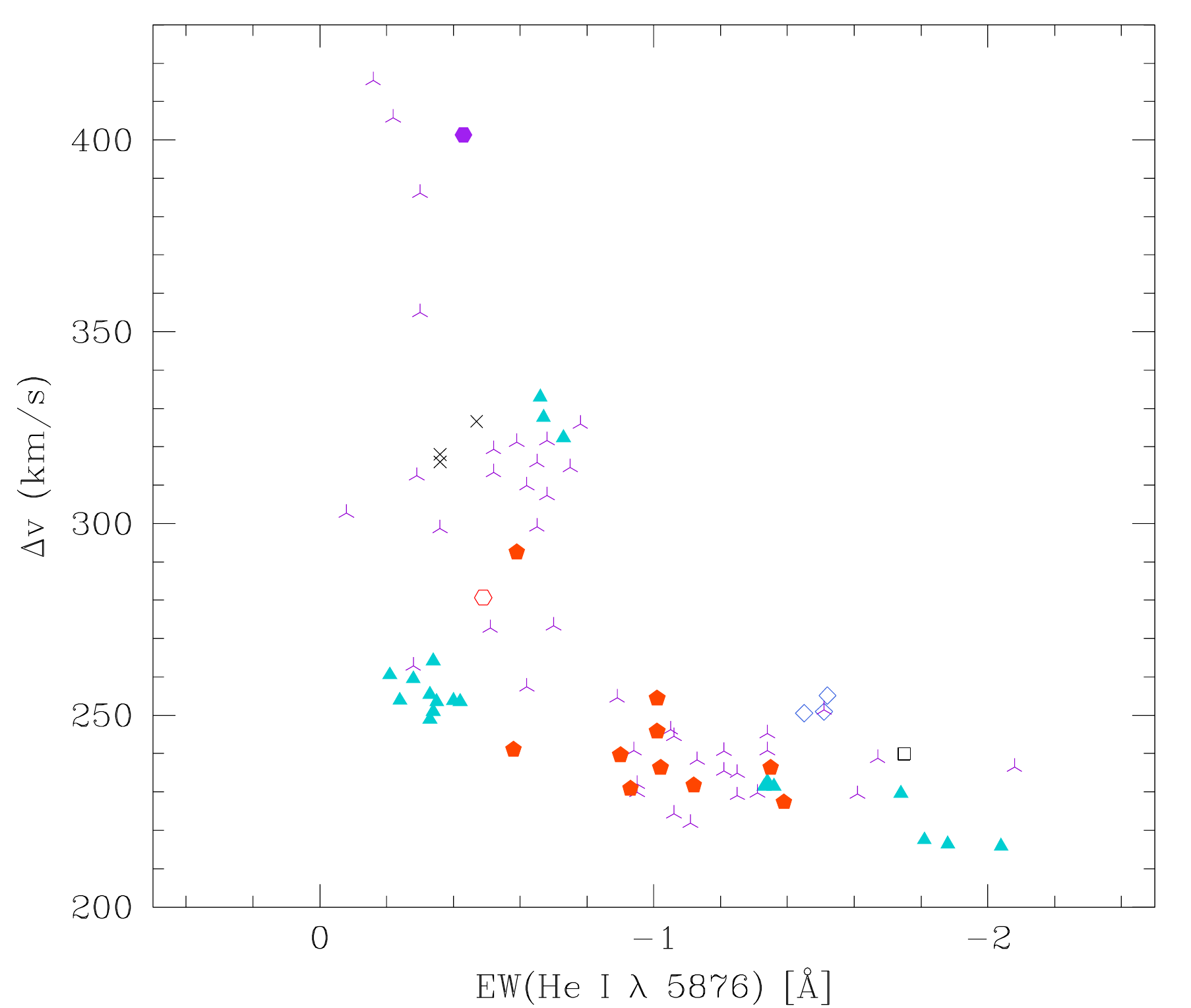}}
    \end{center}
  \end{minipage}
\caption{Relation between the EWs and $\Delta v$ of the double-peaked H$\alpha$ (left), H$\beta$ (middle) and He\,{\sc i} $\lambda$\,5876 emission lines. The symbols have the same meaning as in Fig.\,\ref{EWs}. The dotted line in the left panel illustrates a relation $\Delta\,v \propto |EW(H\alpha)|^{-\gamma/2}$ with $\gamma = 1.184$ as found by \citet{Zam19} for X~Per.}
\label{EWvsdv}
\end{figure*}

Another diagnostic of the disk extension is provided by the equivalent width of the H$\alpha$ emission \citep{Gru06,Zam19}:  
\begin{equation}
  R_{H\alpha} \propto R_*\,|EW(H\alpha)|^{\gamma}
  \label{eq2}
\end{equation}
For X~Per, \citet{Zam19} propose $\gamma = 1.184$. Combining Eqs.\,\ref{eq1} and \ref{eq2}, one would thus expect a relation of the kind
\begin{equation}
  \Delta\,v \propto |EW(H\alpha)|^{-\gamma/2}
  \label{eq3}
\end{equation}
Figure\,\ref{EWvsdv} displays the observed relation between EW and $\Delta\,v$ for the three main emission lines of HD~60\,848. To avoid any bias, we restrict ourselves to those data with a resolving power of 8000 or more. For the H$\alpha$ line, there is considerable scatter. Yet, the relation for the H$\alpha$ line is in qualitative agreement with Eq.\,\ref{eq3}, though it hints at a lower value of $\gamma$ than found for X~Per by \citet{Zam19} and possibly suggests a saturation of $\Delta\,v$ near $\sim 150$\,km\,s$^{-1}$ for the strongest emission states.

During the rising part of a cycle of HD~60\,848, when the H$\alpha$ emission level increases from minimum towards maximum, the corresponding $\Delta\,v$ decreases and reaches its minimum before the emission strength is maximum. Once the decay sets in, $\Delta\,v$ remains low at first, as indicated by the two fast decay events monitored in 2009 and 2019, whilst the EW already decreases. It thus seems that there is a small delay between the variations of these two parameters for H$\alpha$. This could indicate that during disk growth, the disk first expands to reach a maximum extension and then increases its density. Likewise when the decay sets in, the disk maintains its maximum radius for a while, whereas material is already being removed, possibly via a radiation-driven disk-wind or via re-accretion onto the star, from the innermost parts of the disk. 

A somewhat different behaviour is observed for the H$\beta$ line which displays a nearly linear relation between EW and $\Delta\,v$ (Fig.\,\ref{EWvsdv}).  Finally, in the case of the He {\sc i} $\lambda$\,5876 line, the behaviour is clearly different. Two regimes seem to exist. For EW $> -1$\,\AA, the $\Delta\,v$ values range between $250$ and $\sim 400$\,km\,s$^{-1}$ with no clear dependence on EW. On the other hand, Fig\,\ref{EWvsdv} suggests a nearly constant $\Delta\,v$ around $\sim 230$\,km\,s$^{-1}$ for EW $\leq -1.0$\,\AA.

\begin{figure*}
  \begin{minipage}{5.8cm}
    \begin{center}
      \resizebox{5.8cm}{!}{\includegraphics{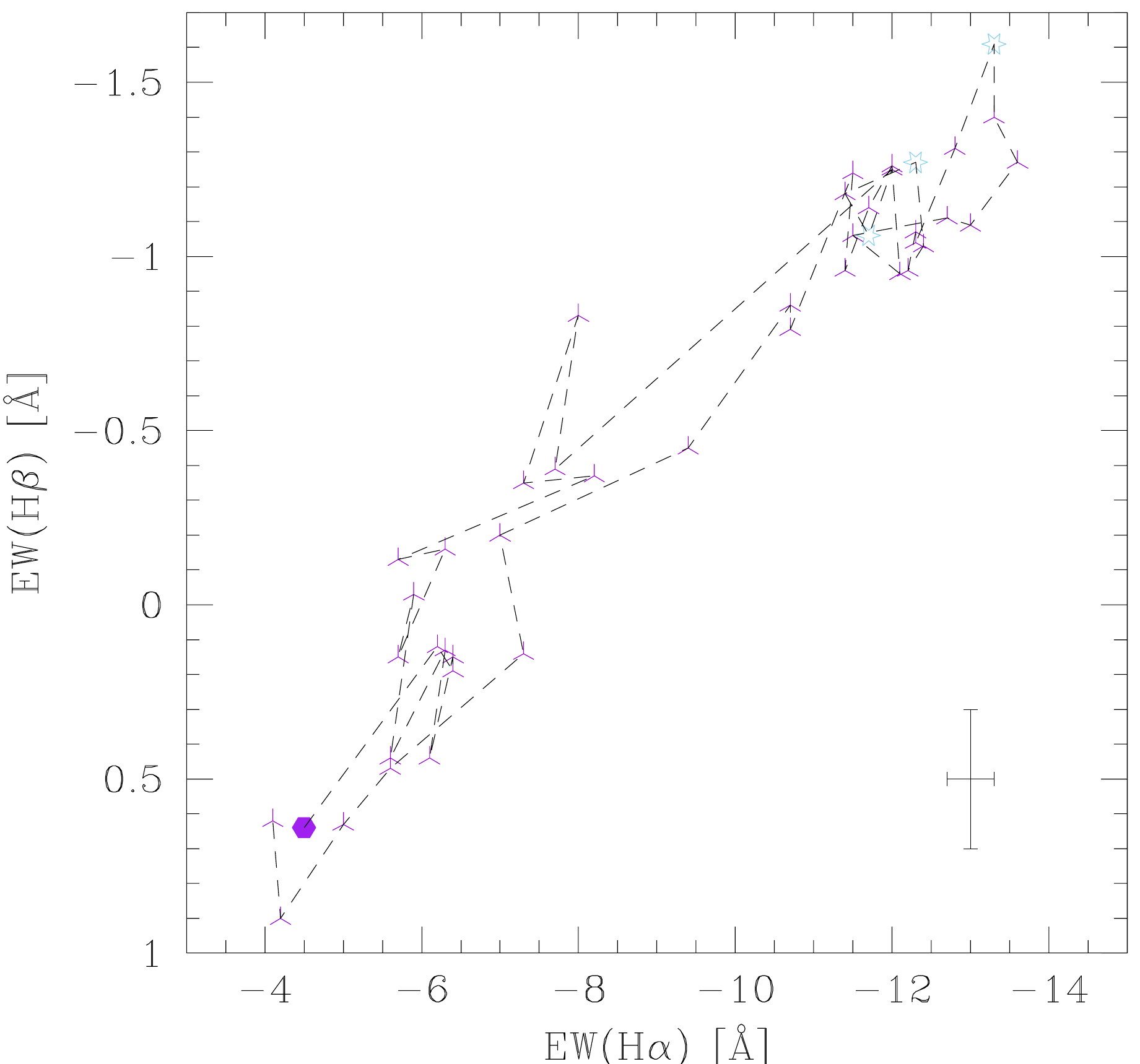}}
    \end{center}
  \end{minipage}
  \hfill
  \begin{minipage}{5.8cm}
    \begin{center}
      \resizebox{5.8cm}{!}{\includegraphics{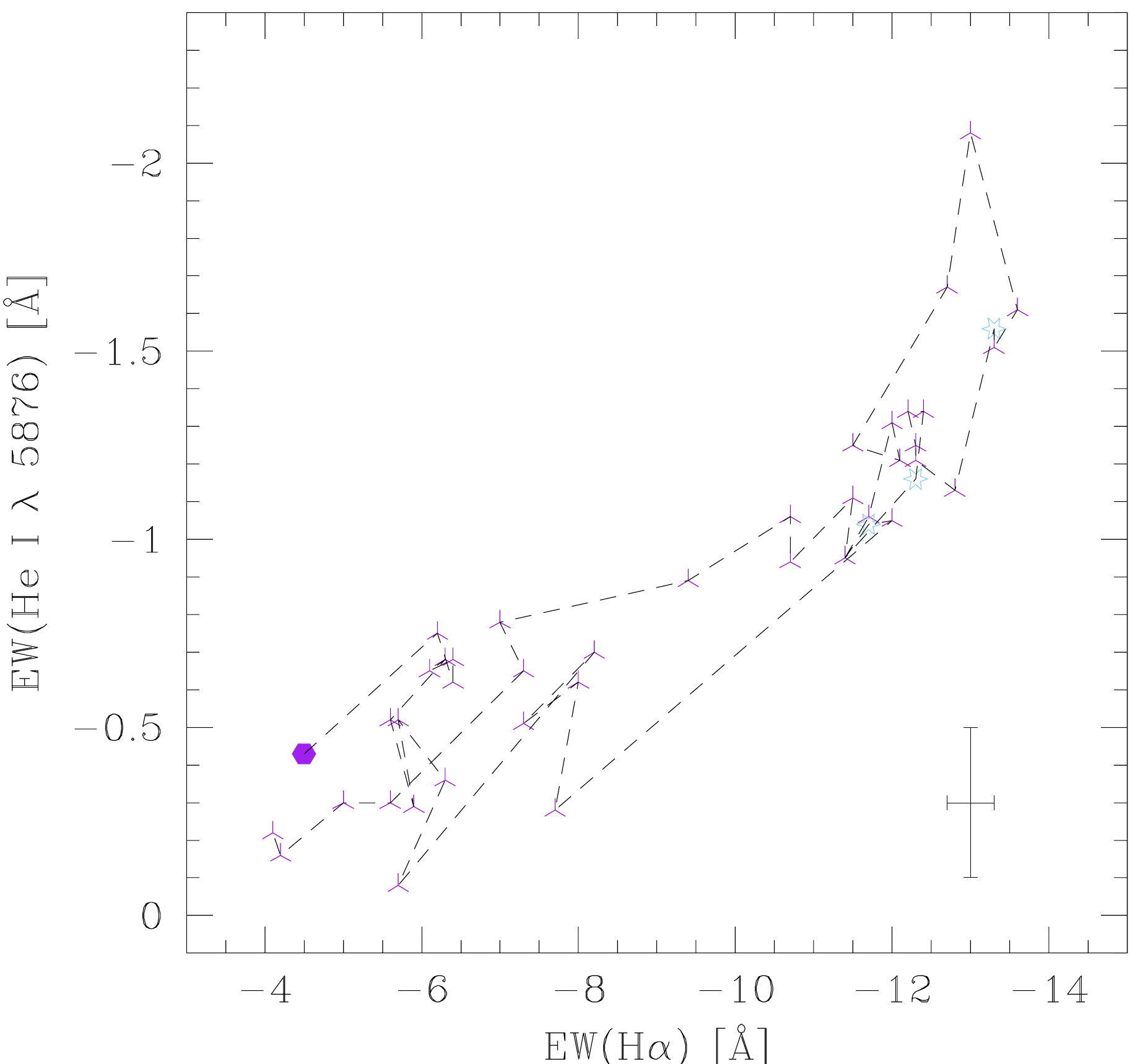}}
    \end{center}
  \end{minipage}
  \begin{minipage}{5.8cm}
    \begin{center}
      \resizebox{5.8cm}{!}{\includegraphics{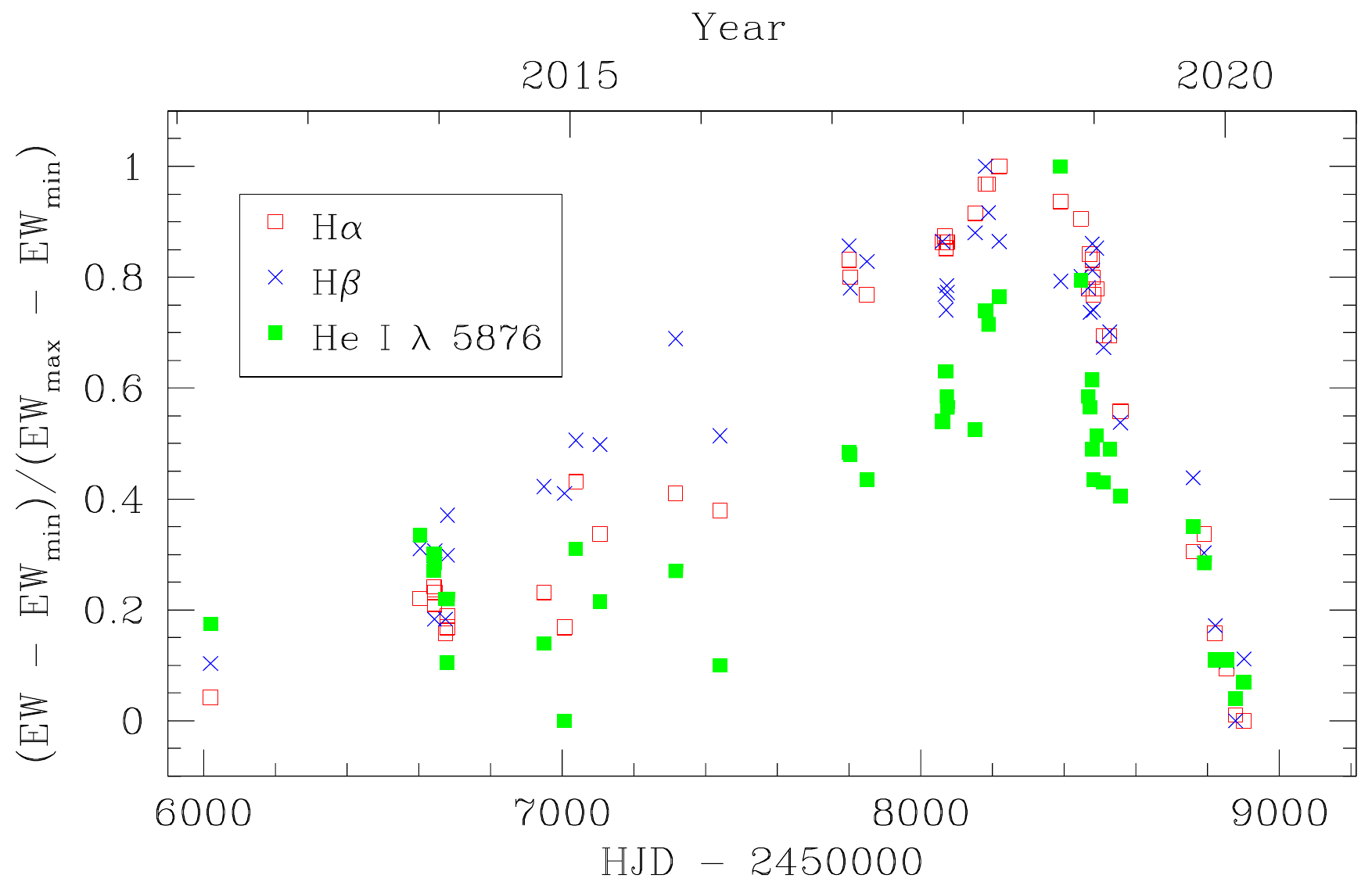}}
    \end{center}
  \end{minipage}
\caption{Relation between the EWs of H$\beta$ and H$\alpha$ (left panel), He\,{\sc i} $\lambda$\,5876 and H$\alpha$ (middle panel), and overall comparison of the normalized EW temporal variations of these lines (right panel) during the most recent cycle of HD~60\,848. The symbols in the left and middle panels have the same meaning as in Fig.\,\ref{EWs}. The typical error bars on the EWs are shown in the bottom right corners of these panels.}
\label{timelag}
\end{figure*}
Interpreting the peak velocity separations as reflecting the characteristic radius of the emission region, we find that the He\,{\sc i} $\lambda$\,5876 emission probes regions that are closer to the star than the H$\beta$ emission which in turn arises from a less extended region than H$\alpha$. In this context, \citet{Rau15} pointed out a possible time lag in the variations of the H$\alpha$, H$\beta$ and He\,{\sc i} $\lambda$\,5876 emission lines: during the 1998 -- 2004  cycle, the apparent strength of the H$\beta$ emission increased first, followed by the H$\alpha$ line and finally by He\,{\sc i} $\lambda$\,5876. At first sight, this behaviour would contradict the expectations for disk dissipation and growth starting in the inner regions close to the star \citep[e.g.][]{Sig13,Lab18}, if the above inferred stratification of the emission regions is correct.

Our new data offer a clearer view of the time delays between the variations of the lines. The left panel of Fig.\,\ref{timelag} reveals that globally the relation between the EWs of H$\alpha$ and H$\beta$ is rather well described by a simple proportionality. Nevertheless, we note that during the early phases of the cycle, the strength of the H$\beta$ emission increases faster than that of H$\alpha$ (see the right panel of Fig.\,\ref{timelag}). An interesting feature are the fluctuations in EW of both lines that produce apparent loops in the left panel of Fig.\,\ref{timelag}. The middle and right panels of Fig.\,\ref{timelag} confirm the existence of a time lag between the variations of the Balmer lines and He\,{\sc i} $\lambda$\,5876. The He\,{\sc i} emission reaches its maximum strength later than the Balmer lines and also displays a narrower maximum. As for the Balmer lines, we observe strong fluctuations of the EW of the He\,{\sc i} line, especially during the early phases of the cycle. In general, the cleanest correlations between the variations of all three lines are observed during the decay phase. The observations of the most recent cycle thus qualitatively confirm the behaviour observed during the 1998 -- 2004 cycle of HD~60\,848. A possible explantion for the behaviour of the He\,{\sc i} $\lambda$\,5876 line could be that the disk first grows in radial extension and then in density. Since the formation of the He\,{\sc i} $\lambda$\,5876 line requires higher excitation conditions, this line most probably forms over the inner regions of the disk (as confirmed by the $\Delta\,v$ values). In our scenario, the radial extension of the He\,{\sc i} emission region would reach a value near its maximum, whilst the emission regions of H$\alpha$ and H$\beta$ are still growing, leading to a stagnation of the EW of He\,{\sc i} $\lambda$\,5876. The subsequent increase of the disk density, once the disk has reached its maximum size, would then correspond to the more rapid growth phase (at near constant $\Delta\,v$) of the EW of He\,{\sc i} $\lambda$\,5876.

Our observations cover three cycles of the disk of HD~60\,848 with maximum H$\alpha$ emission strengths that occurred in early 2001, late 2008 and late 2018.  The variations of the EW over a given cycle are highly asymmetric. At first the emission increases very slowly on timescales of $\sim$ 4\,yr. This is followed by a faster rise to maximum (on a timescale of about 2\,yr). Finally, the decay occurs much faster, on timescales of several months to about half a year. The densest time series at hand is clearly our series of EW(H$\alpha$) measurements consisting of 180 data points. We used the Fourier periodogram method adapted to time series with uneven sampling as developed by \citet{HMM} and \citet{Gos01} to search for a periodicity. The periodogram exhibits no clearly outstanding frequency. The strongest peaks are found at $0.00032 \pm 0.00001$\,d$^{-1}$ and $0.00295 \pm 0.00001$\,d$^{-1}$ with amplitudes near 2.2\,\AA\ and 1.9\,\AA, respectively. These peaks are very likely yearly aliases of each other. The associated timescales are 3125\,d (i.e.\ 8.6\,yr) and 338\,d (i.e.\ 0.93\,yr). However, these timescales do not reflect genuine periodicities as the associated peaks do not clearly stand out in the periodograms and their amplitude are much smaller than the amplitudes of the observed variations. Indeed, whilst the first and second maximum were separated by about 7.6 years, the time interval between the second and third maximum was longer, about 10 years. We thus conclude that whilst the emission strength displays cyclic variations, there is no strict periodicity in this variability. 

\begin{figure}
    \begin{center}
      \resizebox{8cm}{!}{\includegraphics{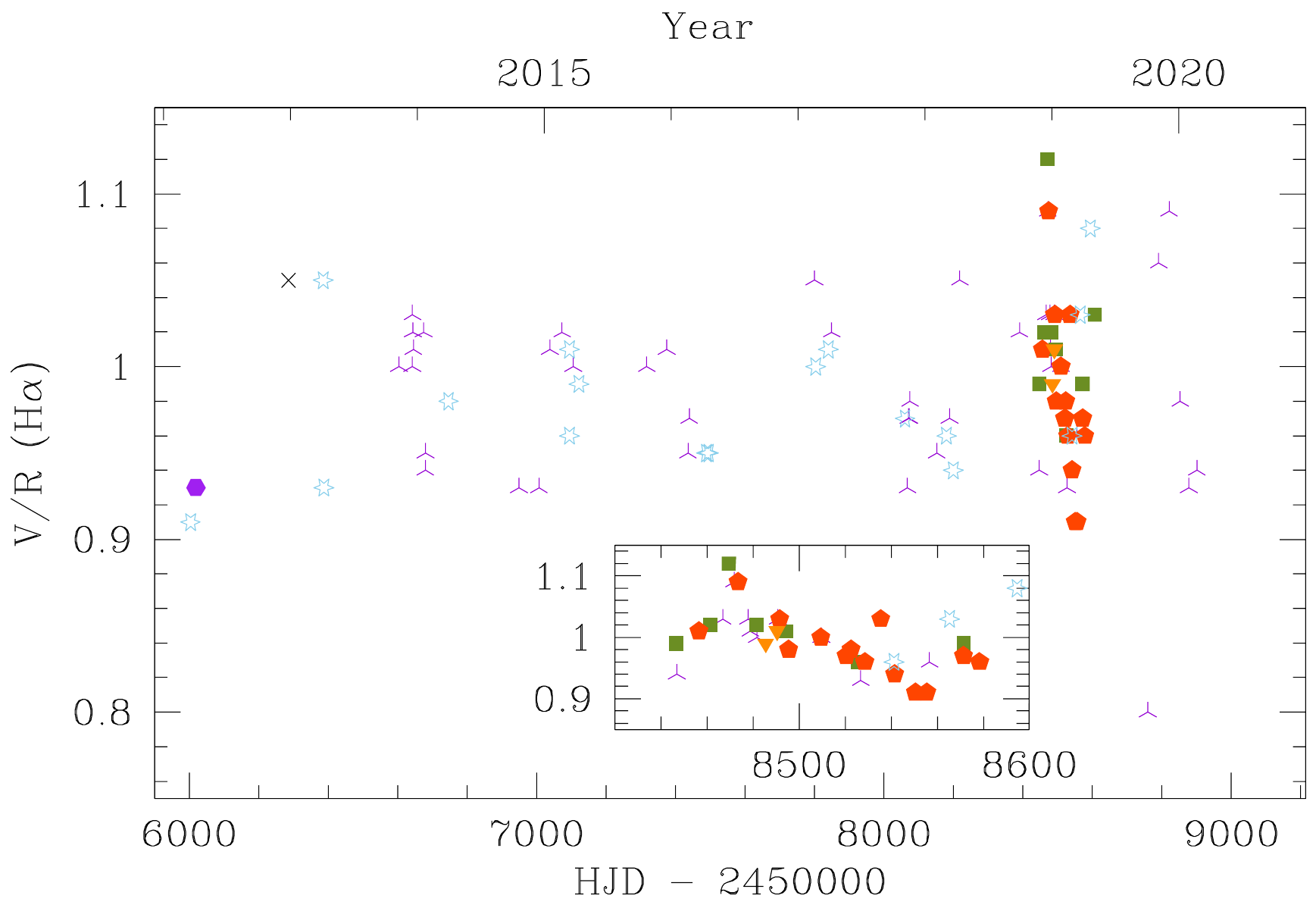}}
    \end{center}
\caption{Variations of the V/R ratio of the H$\alpha$ emission line over the most recent cycle of HD~60\,848. The symbols have the same meaning as in Fig.\,\ref{EWs}. The insert provides a zoom on the 2018 - 2019 decay phase.}
\label{VR}
\end{figure}

The low-amplitude V/R variations of the emissions occur over significantly shorter timescales (Fig.\,\ref{VR}). For the H$\alpha$ line, the periodogram of the V/R ratio exhibits a series of peaks of nearly identical amplitude over the frequency range 0.010 -- 0.100\,d$^{-1}$. Hence, we conclude that the V/R ratio of the H$\alpha$ line varies on timescales of a week to about a hundred days. No dominant timescale can be identified for the V/R variations of the H$\beta$ and He\,{\sc i} $\lambda$\,5876 lines, as their periodograms are consistent with white noise without any dominant peak.

\begin{figure}
    \begin{center}
      \resizebox{8cm}{!}{\includegraphics{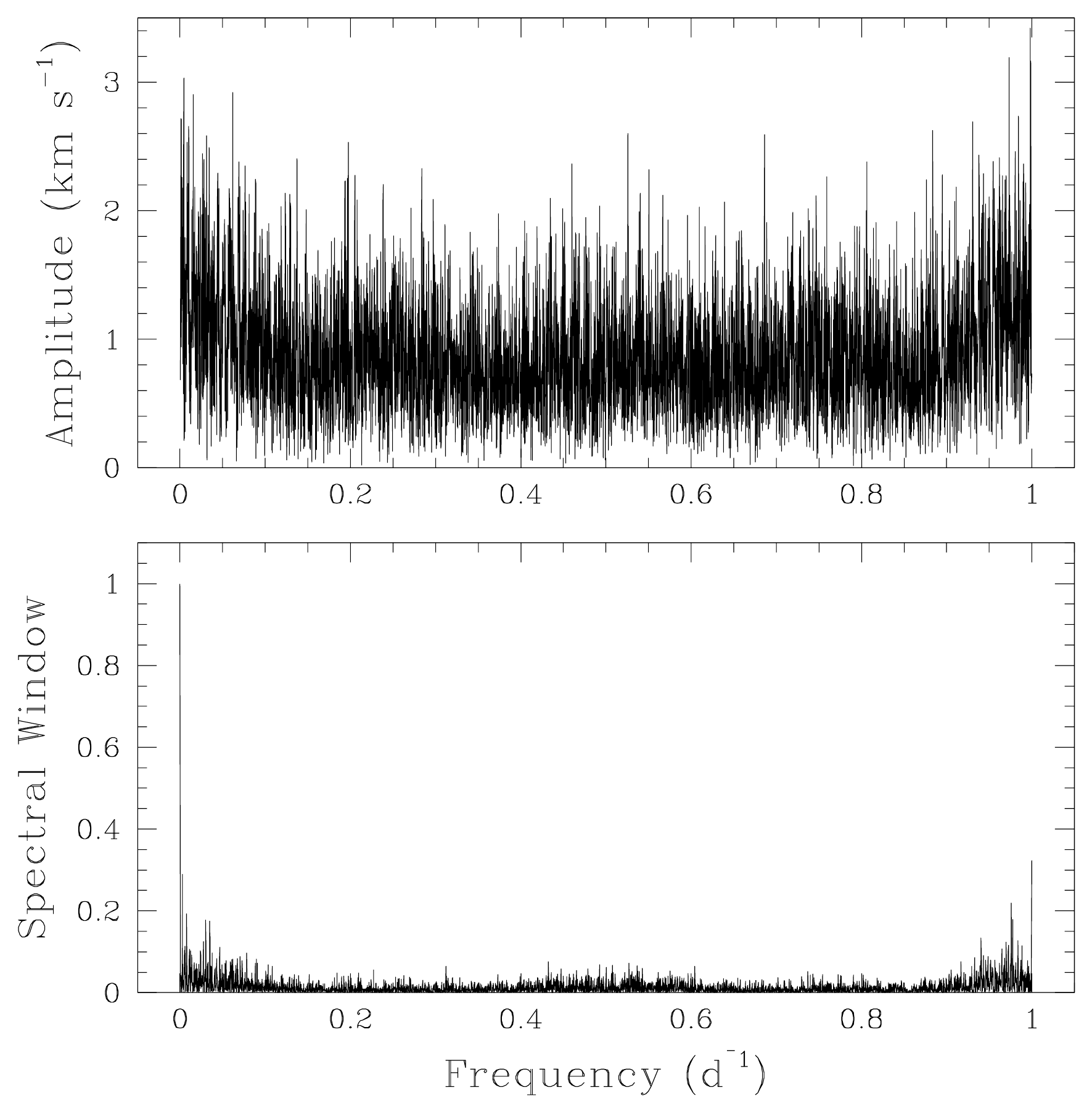}}
    \end{center}
\caption{Top: Fourier periodogram of the RVs of the H$\alpha$ line as determined from its first moment. Only data with a spectral resolving power of more than 8000 were considered here. Bottom: corresponding spectral window.}
\label{Fourier}
\end{figure}

To search for radial velocity (RV) variations of the H$\alpha$ line, which could reflect orbital motion in a binary system, we have computed the moments of the line over the radial velocity interval from $-500$ to $+500$\,km\,s$^{-1}$. The first moment yields the centroid of the line and was taken as a proxy of the RV \citep[e.g.][]{Naz19}. Discarding the MONS data that were found to suffer from larger wavelength calibration uncertainties, we considered several combinations of data depending on their spectral resolution. Considering only data with a resolving power of at least 8000, the mean RV and their dispersion amount to $16.7 \pm 4.7$\,km\,s$^{-1}$. The Fourier analysis of these RVs did not reveal any clear peak with an amplitude exceeding 3\,km\,s$^{-1}$ (see Fig.\,\ref{Fourier}).

\begin{figure}
    \begin{center}
      \resizebox{8cm}{!}{\includegraphics{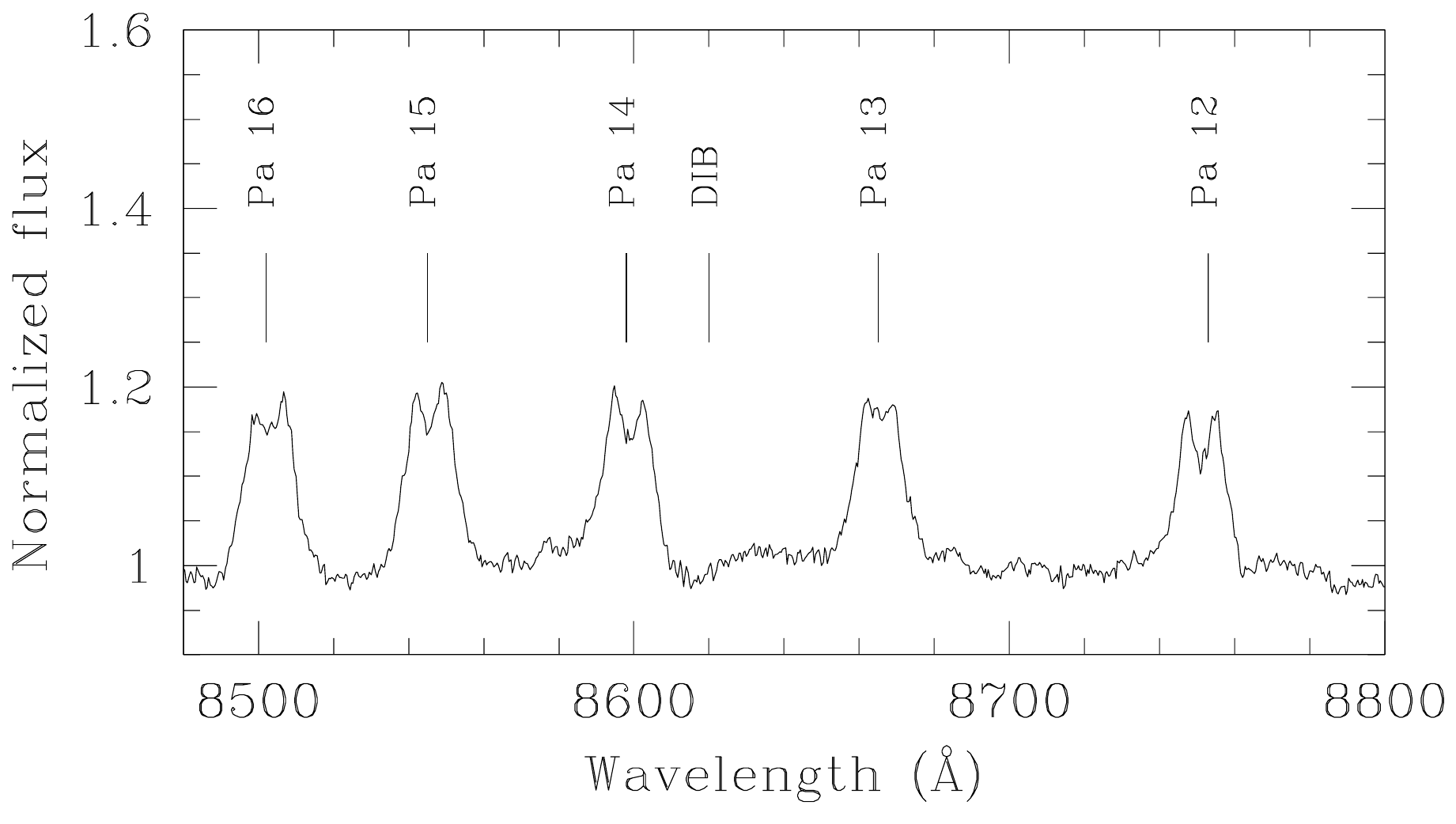}}
    \end{center}
\caption{April 2006 Aur\'elie spectrum of HD~60\,848 in the spectral domain of the Paschen lines.}
\label{GaiaRVS}
\end{figure}

\begin{figure}
    \begin{center}
      \resizebox{8cm}{!}{\includegraphics{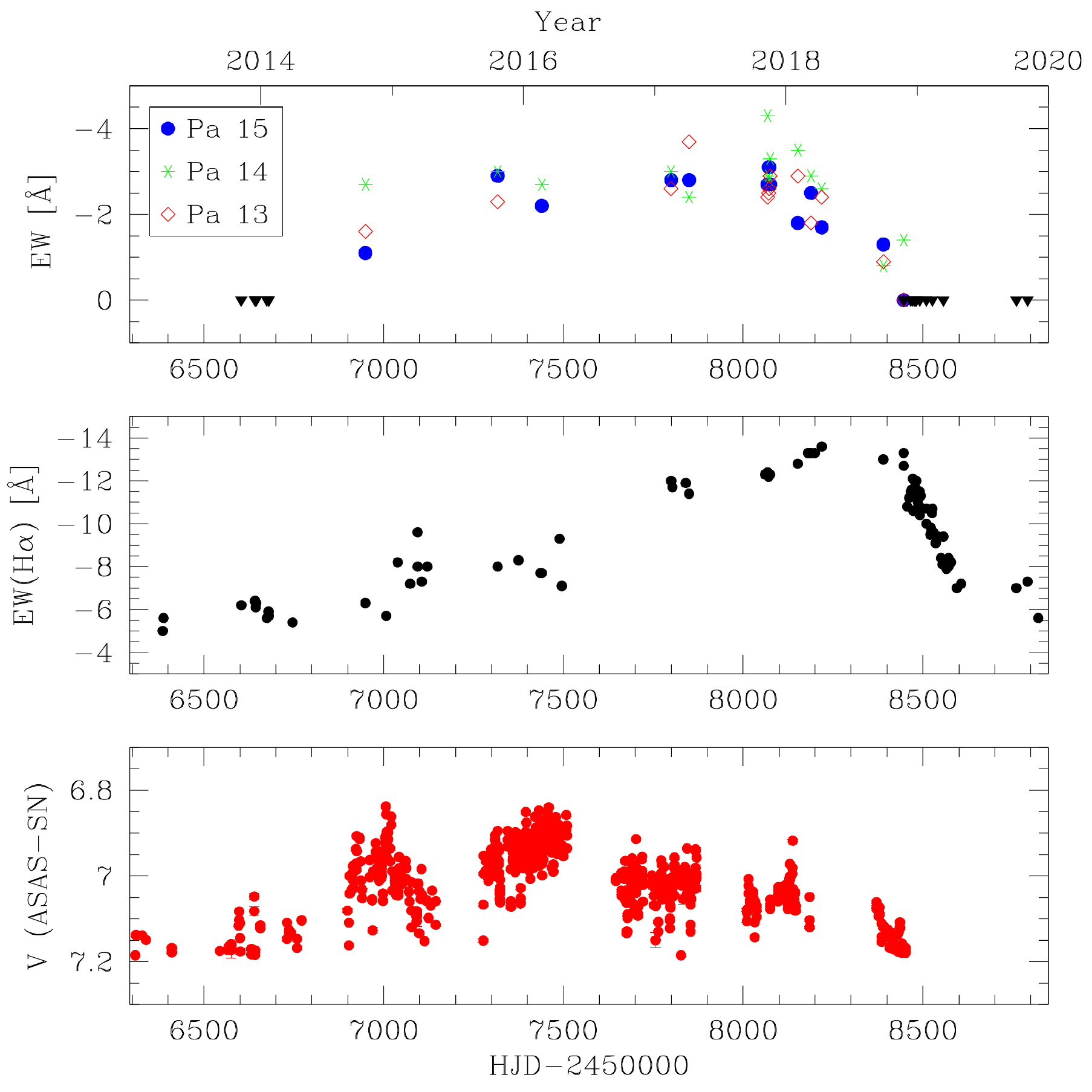}}
    \end{center}
\caption{Top panel: variations of the EWs of the Pa\,13, 14 and 15 lines in the spectrum of HD~60\,848 over the most recent cycle. Downwards pointing triangles indicate dates where the lines were not present in the spectra (neither in emission nor absorption). For comparison, the middle and bottom panels illustrate respectively the variations of EW(H$\alpha$) and $V$ magnitude (from the MAD filtered ASAS-SN data, see Sect.\,\ref{varphotom}) over the same timeframe.}
\label{Paschen}
\end{figure}

In addition to the Balmer lines, double-peaked emissions are also seen in the higher Paschen lines, corresponding to transitions from energy level $n_{\rm up}$ to level 3. These lines were clearly present in the April 2006 OHP spectrum (see Fig.\,\ref{GaiaRVS}), and our TIGRE/HEROS spectra reveal such emissions for Paschen lines corresponding to principal quantum numbers $n_{\rm up}$ between 12 and 20. We have measured the EWs of three of these lines (Pa\,13, 14 and 15, i.e.\ H\,{\sc i} $\lambda\lambda$\,8665, 8598 and 8545). These latter lines are relatively free from blends and from telluric absorptions. Moreover, contrary to Pa\,12, they are not affected by normalization uncertainties due to the proximity of the upper limit of the wavelength domain covered by the HEROS spectrograph. The behaviour of these lines during the 2013 -- 2019 cycle of HD~60\,848 is displayed in Fig.\,\ref{Paschen} and compared to the variations of EW(H$\alpha$) and of the MAD filtered ASAS-SN $V$-band photometry. Several interesting characteristics emerge from this figure. First, the Paschen emission lines are not always present. Spectra from the winter 2013 -- 2014 observing season and from mid-December 2018 onwards reveal neither emission nor absorption in these lines. The April 2006 OHP spectrum reveals EWs near $-2.7$\,\AA\ for all three lines, equal to the strongest emission values seen in our TIGRE/HEROS spectra.
Second, the EW variations of the Paschen lines are more symmetric and do not correlate with those of H$\alpha$, but are in fact quite similar to the photometric variations seen in the ASAS-SN data. In the 2013 -- 2019 cycle, the first detection of the Paschen emission lines occured in October 2014, when the H$\alpha$ line was still at a relatively low emission level of EW(H$\alpha$) = $-6.3$\,\AA. Between October 2015 and November 2017, the strength of the Paschen lines was quite stable at a value close to its maximum, whilst EW(H$\alpha$) changed by a factor $\sim 1.5$ over this time interval. The decay of the Paschen lines set on about one year before that of the H$\alpha$ line. Actually the Paschen lines started to decay whilst H$\alpha$ was still increasing its strength and they reached the detection limit whilst EW(H$\alpha$) was still equal to $-12.7$\,\AA. The $\Delta\,v$ values were $\sim 315$\,km\,s$^{-1}$ when the lines first appeared in October 2014. After this date, they fluctuated around a mean value of 260\,km\,s$^{-1}$ with no well-defined trend. Assuming that these values reflect the location of the emission region \citep{Zam19}, we find that the Paschen lines arise over a similar region as He\,{\sc i} $\lambda$\,5876, although the latter line displays a different behaviour as far as the EWs are concerned. We note that the double-peaked morphology of the Paschen lines is less obvious than for the other emission lines discussed above, and we sometimes observe either a single peak or more than two peaks for a given line. Finally, we stress that, although Fig.\,\ref{Paschen} focuses on the EWs of three out of the nine Paschen lines, all nine lines display a similar behaviour.

The atomic level populations in the circumstellar disk of Oe/Be stars result from statistical equilibrium under the effect of the photoionizing radiation from the central star \citep[e.g.][]{SigJon}. The difference in behaviour of the Paschen lines and the H$\alpha$ line most likely stems from the fact that the disk is optically thin in the Paschen lines, but optically thick in H$\alpha$. Indeed, the $\log{(gf)}$ values of these lines differ by more than 2.0\,dex. 

\begin{figure}
    \begin{center}
      \resizebox{8cm}{!}{\includegraphics{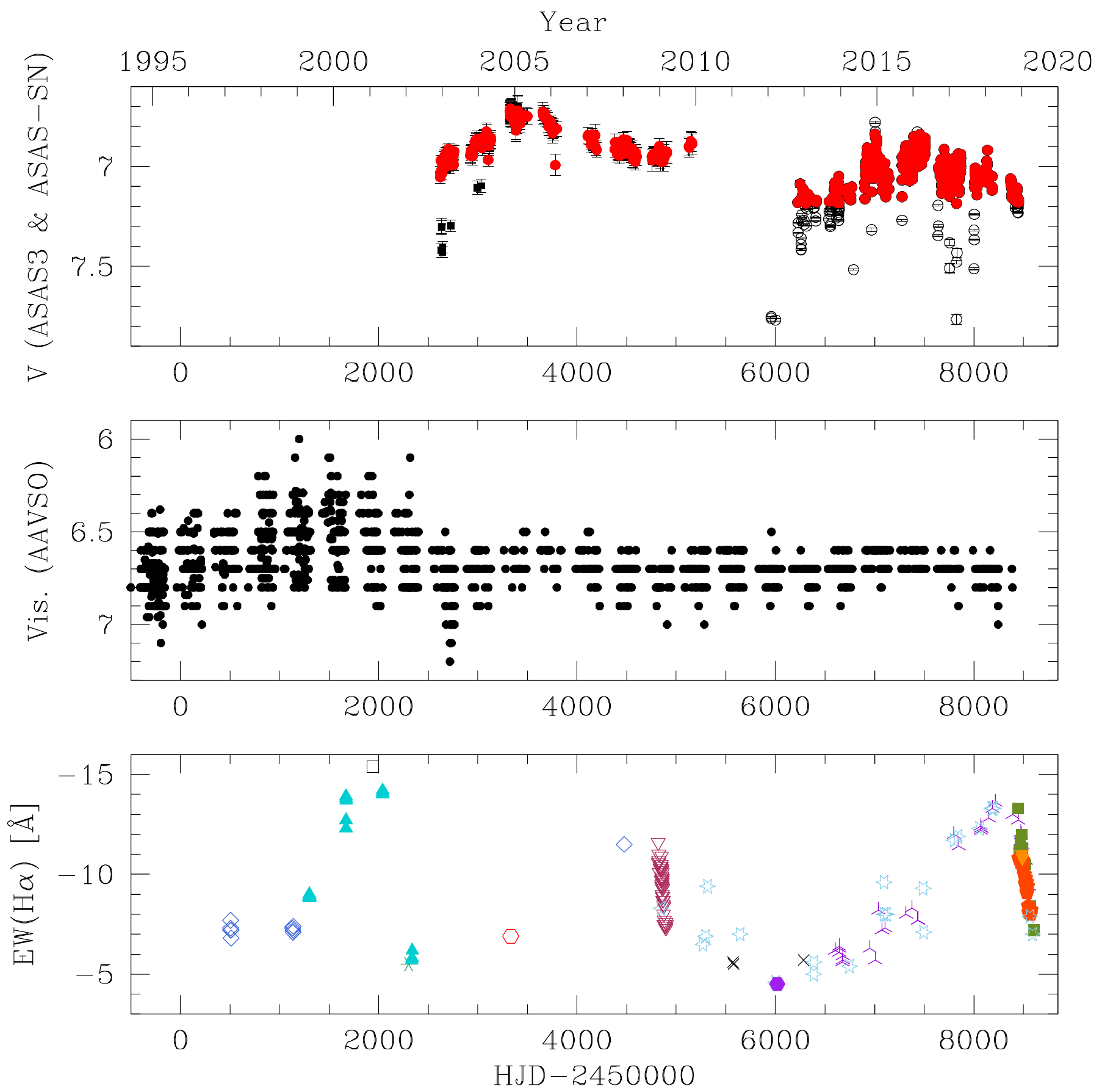}}
    \end{center}
\caption{Photometric measurements of HD~60\,848 along with the variations of the EW of H$\alpha$. In the ASAS panel, squares and circles stand for ASAS-3 and ASAS-SN data respectively. The empty circles indicate the original time series, whereas the filled red symbols correspond to the data after MAD filtering.}
\label{photometry}
\end{figure}

\citet{Cho17} presented six H-band spectra of HD~60\,848 taken between 19 January 2014 and 16 April 2014, i.e.\ during the early phases of the increase of the emission strength. The EW of the Br\,11 emission varied between $-4.8$ and $-6.1$\,\AA, and the $\Delta\,v$ values ranged between 287 and 318\,km\,s$^{-1}$. Beside some increasing trend of the line strength, there were also short-term fluctuations similar to what we report here. \citet{Lab18} argue that the Br\,11 emission arises from regions between about 2 and 6\,R$_*$. At first sight, the presence of a relatively strong Br\,11 emission at this epoch is somewhat puzzling in view of the simultaneous lack of Paschen lines. Indeed, the Paschen lines made their first appearance about six months later. However, this is probably again an issue of optical depth as the $\log{(gf)}$ of Br\,11 is more than 0.7\,dex larger than those of the Paschen lines discussed above. 

\subsection{Photometric variability} \label{varphotom}
Photometric variations are common among Be stars \citep[e.g.][and references therein]{Mou98,Sem18}. Non-radial pulsations and/or rotational modulations of spotted stellar surfaces are thought to be responsible for short-period (less than a day to a few days) variations with amplitudes of several hundredths of a magnitude. Binary Be stars display similar amplitude variations on longer time-scales up to hundreds of days \citep{Baa94}. Finally, quasi-cyclic or irregular variations occur on timescales of years or decades with amplitudes up to several tenths of a magnitude. In the case of HD~60\,848, apparently irregular photometric variability was reported for the first time by \citet{Hof34}. 

\begin{figure*}[t!]
    \begin{center}
      \resizebox{!}{5.2cm}{\includegraphics{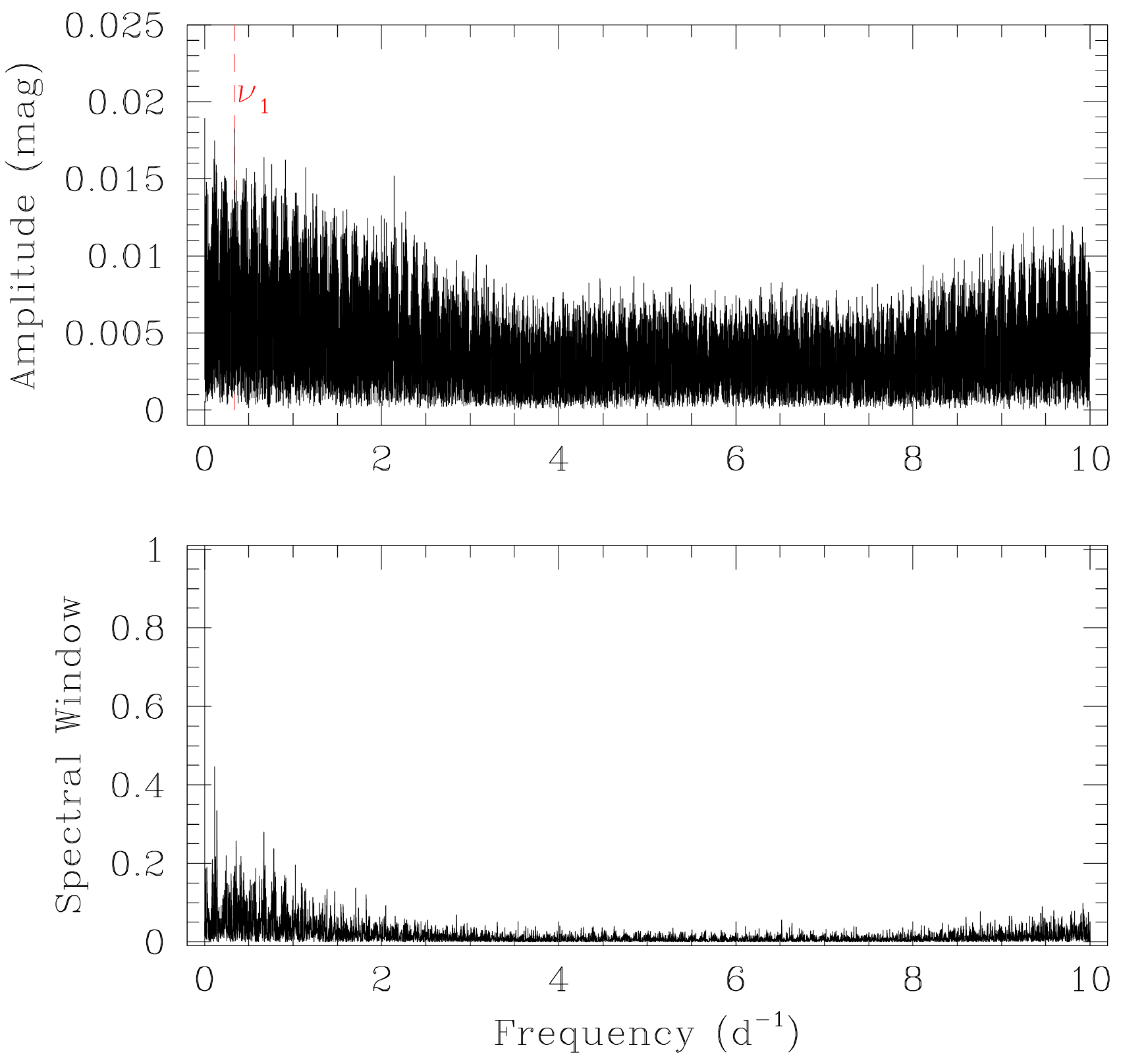}}
      \resizebox{!}{5.2cm}{\includegraphics{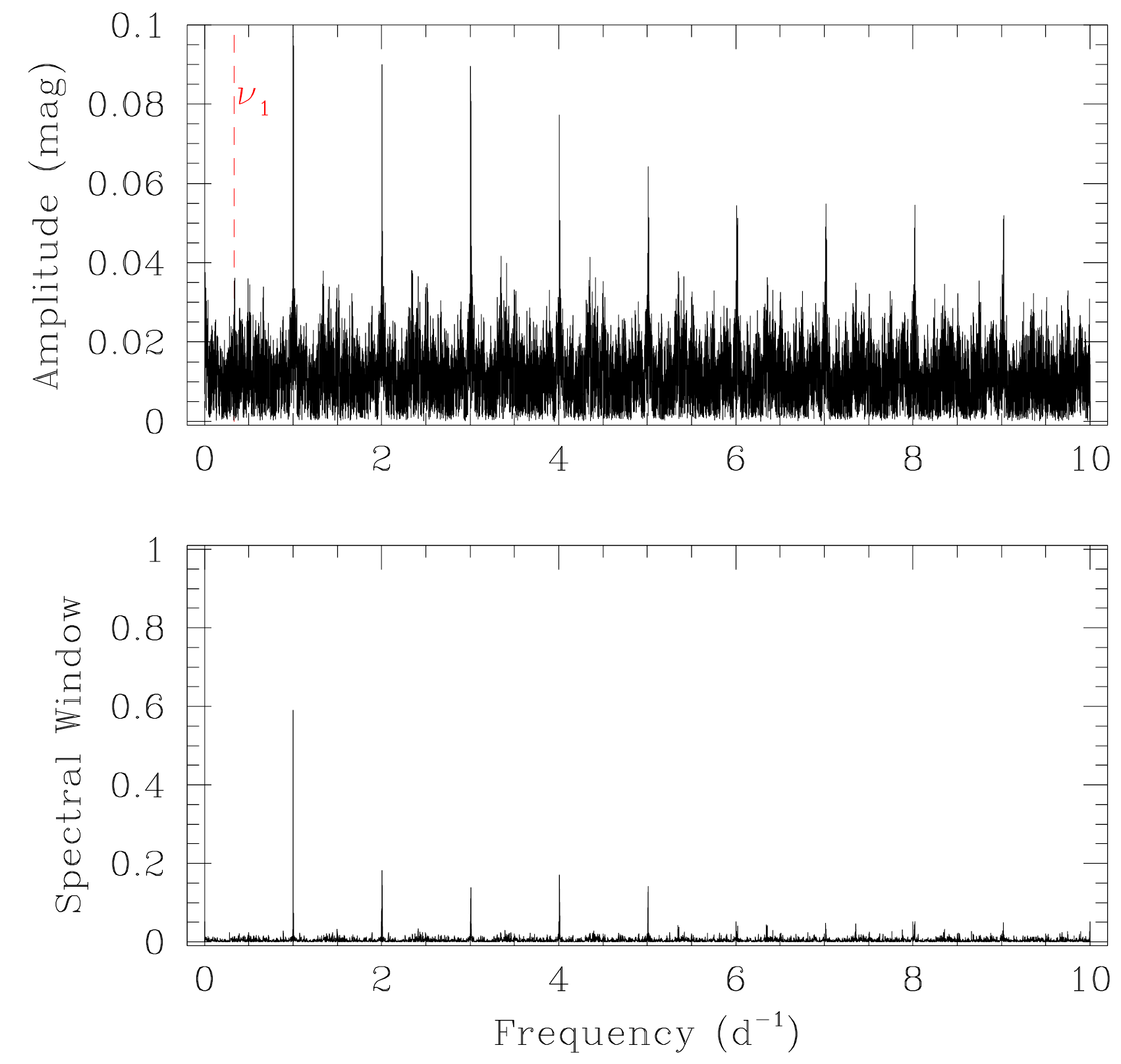}}
      \resizebox{!}{5.2cm}{\includegraphics{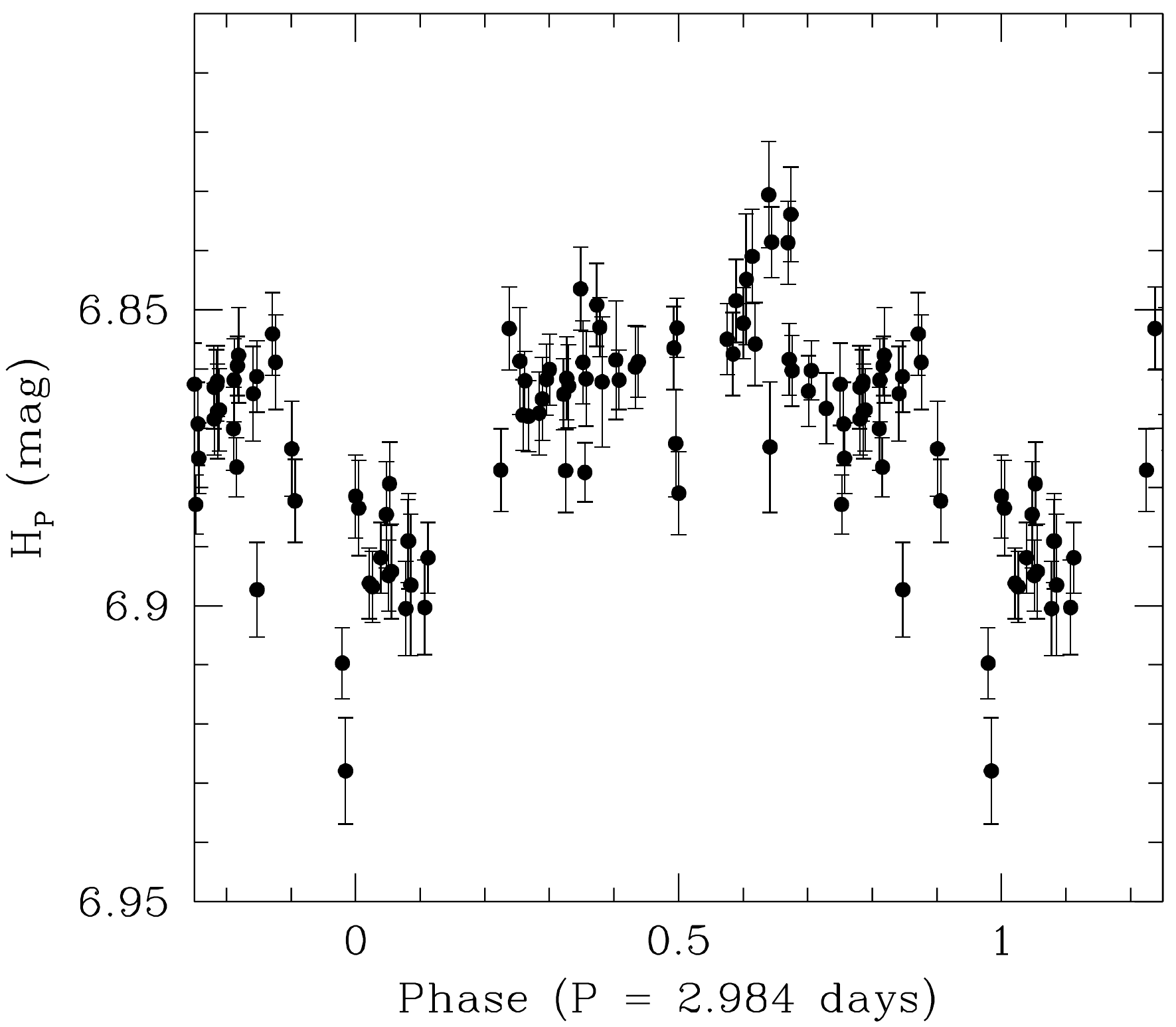}}
    \end{center}
    \caption{{\it Left}: periodogram (top) and spectral window (bottom) of the {\it Hipparcos} photometry. The dashed red line indicates the position of the $\nu_1$ frequency. {\it Middle}: same for the ASAS-3 data. {\it Right}: {\it Hipparcos} photometry folded with the 2.984\,d period.}
\label{cycleHip} 
\end{figure*}
Figure\,\ref{photometry} illustrates the available photometric data. The longest time series is provided by the AAVSO visual magnitudes. Whilst these data suggest considerable photometric variability (over the range from 7.3 to 5.9 in visual magnitude, with a smaller range from 7.2 to 6.4\,mag over the last eight years), they do not provide error estimates rendering a quantitative assessment of the significance of the variability difficult. Considering the variations over the last 25 years, we find no correlation between the strength of the H$\alpha$ emission and the AAVSO visual magnitudes (see Fig.\,\ref{photometry}). Since the spectroscopic and photometric measurements are generally not simultaneous, we interpolated between the AAVSO measurements taken at dates bracketing those of the H$\alpha$ observations\footnote{We only retained interpolations where the AAVSO data were taken within at most 10\,days of the spectroscopic observation.}. The scatter plot of the visual magnitudes as a function of EW(H$\alpha$) reveals neither a correlation nor an anti-correlation. The Pearson correlation coefficient amounts to $r = -0.32$, whilst the Spearman rank correlation coefficient amounts to $r_s = -0.43$. This finding constrasts with the conclusion of \citet{Mou98} who analysed the long-term (over 40 years) spectrophotometric variations of HD~60\,848 and found the Balmer discontinuity (which was considered as a proxy of the strength of line emission) to be well correlated with the continuum emission. According to \citet{Mou98}, line emission is stronger when the star is brighter and redder. 

To further investigate this issue, we considered the ASAS-3 and ASAS-SN data. The MAD filtering process applied to the ASAS-SN data removes the data points with very low fluxes (see Fig.\,\ref{photometry}), and might actually cut out some real variations between HJD\,2\,456\,000 and 2\,457\,000. The remaining 730 data points indicate that HD~60\,848 brightened between 2012 and 2015 and faded after 2017. Again, no simple correlation is found with the strength of the H$\alpha$ emission. Contrary to the EW(H$\alpha$) changes, the photometric increase and decay phases display rather similar slopes. Moreover there seems to be a time delay between the maximum optical brightness and the maximum EW(H$\alpha$) but its value differs between the 2008 and 2018 maximum EW(H$\alpha$) events. 

Finally, the {\it Hipparcos} photometry collected between HJD\,2\,447\,967 and  2\,449\,057 indicates variations with a peak-to-peak amplitude of $\Delta H_p = 0.1$\,mag, and a broad brightness maximum around HJD\,2\,448\,600 (i.e.\ December 1991). Unfortunately, there are no contemporaneous spectroscopic data available, preventing us from searching for correlations with the H$\alpha$ emission strength. Yet, the low amplitude of the variations observed with {\it Hipparcos} suggests that the actual photometric variability of HD~60\,848 has amplitudes lower than suggested by the AAVSO data.

\citet{Lef09} reported the detection of a period near 2.983\,d in the {\it Hipparcos} photometry of HD~60\,848. A Fourier periodogram of these data (see Fig.\,\ref{cycleHip}) reveals a very low frequency which reflects long-term variations, as well as a peak at a frequency of $\nu_1 = 0.335$\,d$^{-1}$, corresponding to a period of 2.984\,d in agreement with the results of \citet{Lef09}. When folded with this period, the {\it Hipparcos} photometry reveals a non-sinusoidal variation with a peak-to-peak amplitude of about 0.04\,mag (Fig.\,\ref{cycleHip}). We searched for confirmation of the existence of this period in the MAD-filtered ASAS-3 and ASAS-SN data. These data were collected over longer time intervals and their periodograms are even more dominated by the long-term trends. Whilst the ASAS-SN data reveal no other significant peak in their periodogram, the ASAS-3 data reveal a peak at $\nu_1$ (with an amplitude of 0.035\,mag) as well as at its daily aliases ($1$\,d$^{-1} - \nu_1$, $\nu_1 + 1$\,d$^{-1}$, $2$\,d$^{-1} - \nu_1$, $\nu_1 + 2$\,d$^{-1}$, etc., see Fig.\,\ref{cycleHip}). Actually, the strongest alias is found at $\nu_1 + 4$\,d$^{-1}$ rather than at $\nu_1$ (see also Sect.\,\ref{Disc}). However, both the {\it Hipparcos} and the ASAS-3 datasets do not well sample such high frequencies: except for a handful of data points, the overwhelming majority of the data were acquired at a rate of less than one observation per day. In view of the amplitudes recorded (0.018\,mag in the {\it Hipparcos} data, 0.035\,mag in the ASAS-3 photometry), we find that the visibility of this signal might change with time, a feature which is not uncommon among Be stars \citep[e.g.][]{Hua09,Sem18,Naz20}. Establishing the precise value of the frequency of these short-term photometric variations and constraining their properties will only be possible by means of future high-cadence, high-precision photometric observations. 

\subsection{The X-ray emission}
HD~60\,848 is an isolated field object: it is neither part of a cluster \citep{deW04}, nor embedded in a H\,{\sc ii} region \citep{Rey82}. Our {\it XMM-Newton} and {\it Chandra} X-ray images indeed reveal a rather isolated source which lacks a population of (X-ray bright) pre-main sequence stars that is frequently seen close to massive stars in young open clusters \citep[e.g.][]{Rauw03,Sana}. Moreover, high-angular resolution observations revealed no astrometric companions \citep{Mas09,Ald15}. Therefore, source confusion should not be an issue and the observed X-ray emission must come from HD~60\,848 itself.

Spectral fits were performed in Xspec v12.9.1p, considering solar abundances of \citet{And89}, as was done in \citet{Rau13}. The {\it XMM-Newton} spectrum of April 2012 was fitted by \citet{Rau13} with an absorbed optically-thin thermal plasma model. No absorption was needed beyond that due to the interstellar medium ($3\times 10^{20}$\,cm$^{-2}$), the plasma temperatures amounted to $kT_1 = 0.12$\,keV and $kT_2 = 0.83$\,keV, and the observed flux reached $(3.2 \pm 0.3) \times 10^{-14}$\,erg\,cm$^{-2}$\,s$^{-1}$ in the 0.5 -- 10\,keV energy band. The X-ray to bolometric luminosity ratio was $\log{\frac{L_X}{L_{\rm bol}}} = -7.29 \pm 0.07$, towards the lower end of the range observed for O-type stars \citep{Naz09}.

\begin{figure}
    \begin{center}
      \resizebox{8cm}{!}{\includegraphics{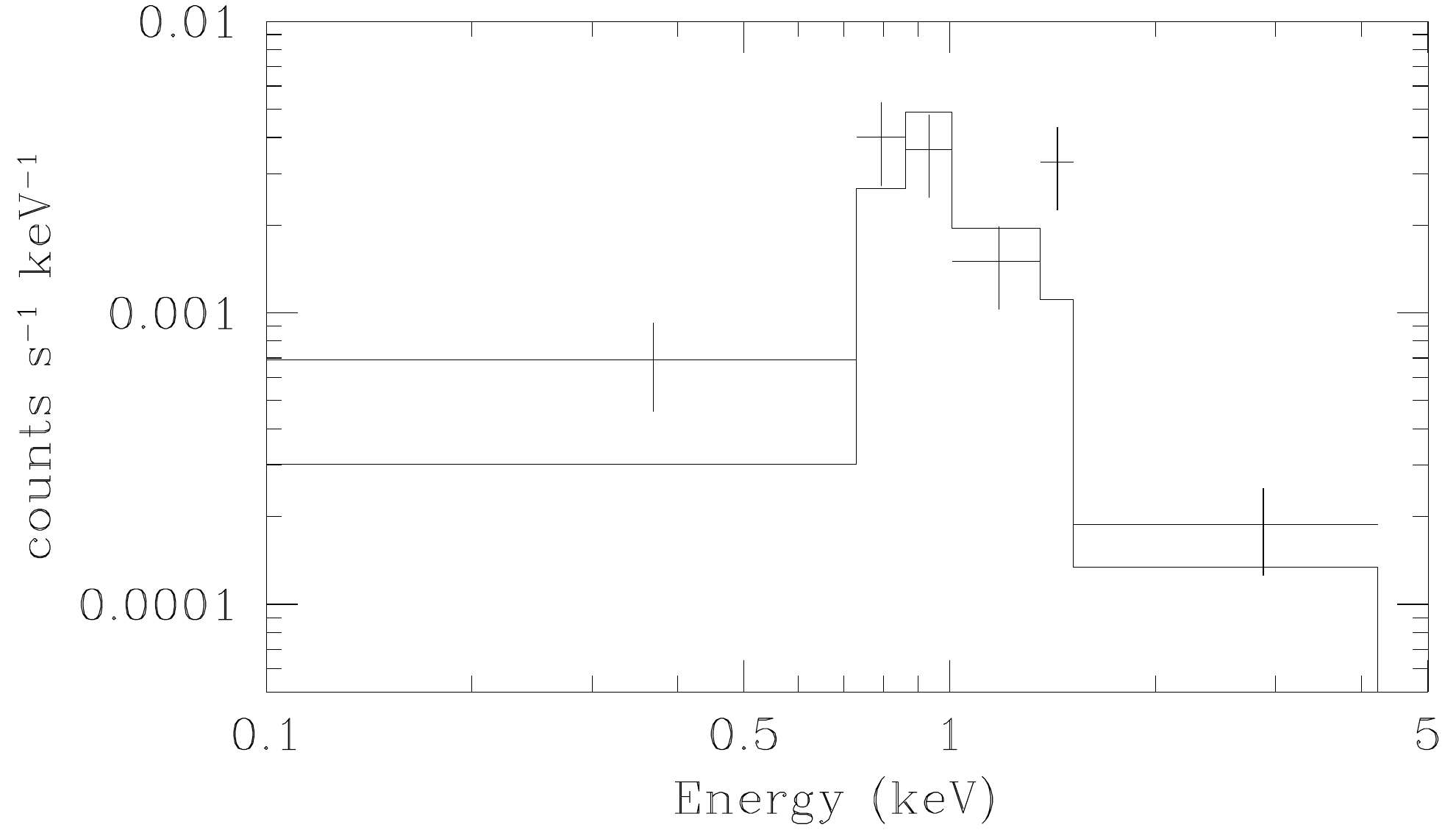}}
    \end{center}
    \caption{{\it Chandra}-ACIS X-ray spectrum of HD~60\,848 taken on 30 December 2018 (data points with error bars). The histogram displays the spectral model of the 2 April 2012 {\it XMM-Newton} EPIC observation scaled to best match the {\it Chandra} data (see text).}
\label{Xrays} 
\end{figure}

The new {\it Chandra} spectrum contains only $\sim 50$ net counts for the source. A spectral fit using an absorbed single-temperature model ({\tt phabs $\times$ apec}) was of poor quality. A better fit quality was obtained by fixing instead the temperatures, absorbing column, and normalization factor ratio to those found for the {\it XMM-Newton} data, and allowing only the overall normalization to vary in the fitting procedure. In this way, we obtained the fit displayed in Fig.\,\ref{Xrays} and derived an observed flux of $(4.3^{+0.7}_{-0.6}) \times 10^{-14}$\,erg\,cm$^{-2}$\,s$^{-1}$. Finally, folding the model of the {\it XMM-Newton} spectrum through the {\it Chandra} response matrices yielded a theoretical count rate of $0.0019$\,cts\,s$^{-1}$ while the actual observed ACIS count rate amounts to $0.0031 \pm 0.0006$\,cts\,s$^{-1}$.

Therefore, HD~60\,848 may be slightly X-ray brighter at the time of the {\it Chandra} observation. However, in view of the errorbars and remaining cross-calibration uncertainties between {\it XMM-Newton} and {\it Chandra}, this brightening does not appear significant. Furthermore, the X-ray spectrum remains soft. There is thus no evidence of a $\gamma$~Cas-like emission, which typically has $kT = 10$\,keV. Our observations rather suggest that the X-ray emission of HD~60\,848 is essentially independent of the properties of its Be disk. In view of the X-ray luminosity level and its soft character, it is very likely that the X-ray emission arises inside a stellar wind, as is usual in O-stars. Indeed, based on {\it IUE} data, \citet{How89} evaluated a mass-loss rate of $\log{\dot{M}} = -6.8$ for the star, where $\dot{M}$ is expressed in M$_{\odot}$\,yr$^{-1}$, and \citet{Pri90} determined a wind velocity of 1720\,km\,s$^{-1}$. HD~60\,848 thus possesses a stellar wind quite typical for late O-type stars, which is perfectly able to generate the observed X-ray emission.

\section{Discussion} \label{Disc}
\subsection{Fundamental parameters}
From the Aur\'elie spectra taken in December 2009, when the H$\alpha$ emission of the disk was low, we derived EW(He\,{\sc i} $\lambda$\,4471) = $0.349 \pm 0.013$\,\AA\ and EW(He\,{\sc ii} $\lambda$\,4542) = $0.276 \pm 0.019$\,\AA. Using the \citet{Con71} criterion, these EWs lead to $\log{W'} = \log{\left(\frac{EW(He\,I \lambda\,4471)}{EW(He\,II \lambda\,4542)}\right)} = 0.102 \pm 0.034$ which corresponds to a spectral type O8 (with O7.5 within the uncertainty). Whilst this result agrees with many previous determinations \citep[e.g.][]{Pla24,Con77,Sot11}, we note that the He\,{\sc i} $\lambda$\,4471 line displayed some traces of residual emission which could bias the classification towards earlier spectral types \citep{Neg04}. As shown by Fig.\,\ref{timelag}, the emission of He\,{\sc i} $\lambda$\,5876 does not reach its minimum at the same time as H$\alpha$. Assuming that the circumstellar He\,{\sc i} $\lambda$\,4471 emission follows the same behaviour as He\,{\sc i} $\lambda$\,5876, we measured EW(He\,{\sc i} $\lambda$\,4471) = $0.455 \pm 0.008$\,\AA\ and EW(He\,{\sc ii} $\lambda$\,4542) = $0.250 \pm 0.015$\,\AA\ on a TIGRE spectrum taken in December 2014 when He\,{\sc i} $\lambda$\,5876 was at its minimum emission strength (though the EW was still negative). These results yield $\log{W'} = 0.260 \pm 0.027$, i.e.\ an O8.5 spectral type with O9 at a bit more than $1\,\sigma$.

Comparing the strength of He\,{\sc ii} absorptions with those in spectra of standard O stars, \citet{Neg04} reclassified HD~60\,848 as an O9.5\,IVe star. We have compared the strength of the observed He\,{\sc ii} $\lambda$\,4542 line with synthetic TLUSTY spectra from the OSTAR2002 grid \citep{Lan03}. This comparison suggests an effective temperature of $(31\,700 \pm 350)$\,K. In view of these results, it seems reasonable to assume that the most likely spectral type of HD~60\,848 is O9 with an uncertainty of half a spectral subclass.

The second {\it Gaia} data release \citep[DR2,][]{GaiaDR2} leads to a distance estimate for HD~60\,848 of $2562^{+609}_{-435}$\,pc \citep{Bai18}. Adopting a mean $V$ magnitude\footnote{The $m_V$ estimate was taken as the mean of the values quoted by \citet{Reed}, whilst the quoted uncertainty corresponds to the range of variations seen in the {\it Hipparcos} lightcurve (Fig.\,\ref{cycleHip}). The $E(B-V)$ colour excess and its uncertainty were estimated as the mean and dispersion of the value obtained by \citet{DipSav} and the colour excess corresponding to the observed colours compiled by \citet{Reed} with respect to the intrinsic $(B-V)_0$ of late O-type stars given by \citet{MartPlez}.} of $m_V = 6.90 \pm 0.10$, $E(B-V) = 0.09 \pm 0.02$, and $R_V = 3.1$, we derive an absolute magnitude of $M_V = -5.42^{+0.40}_{-0.53}$. This result indicates a giant luminosity class. We thus classify HD~60\,848 as an O8.5-9.5\,IIIe star. From there, we would infer a stellar radius of about 14.5\,R$_{\odot}$. However, because of the circumstellar disk, the optical brightness of the star could actually be overestimated. Discarding the most deviating ASAS-3 and ASAS-SN data points, we find that the lowest observed $V$-band flux corresponds to $m_V \simeq 7.2$. Assuming that this value reflects the genuine photospheric emission, we then estimate $M_V \simeq -5.12$ and a smaller stellar radius of about $12.6$\,R$_{\odot}$. 

\citet{Fre05} used a non-LTE model atmosphere code accounting for gravity darkening and rotational flattening to compute synthetic H$\gamma$, He\,{\sc i} $\lambda$\,4471 and Mg\,{\sc ii} $\lambda$\,4481 photospheric absorptions. The synthetic profiles were compared to spectra of B-type stars to infer their fundamental parameters and quantify the impact of rotation on these parameters via the definition of so-called {\it parent non-rotating counterpart} ({\it pnrc}) parameters. For HD~60\,848, \citet{Fre05} derived the {\it pnrc} parameters $T_{\rm eff} = 28\,491 \pm 630$\,K, $\log{g} = 4.06 \pm 0.07$, $v\,\sin{i} = 256 \pm 16$\,km\,s$^{-1}$ and $i = (35.4 \pm 1.8)^{\circ}$. The inclination inferred for the star is in excellent agreement with the disk inclination inferred from the H$\alpha$ profiles \citep[$\sim 30^{\circ}$,][]{Rau15,Rau18}, indicating that the disk is located in the equatorial plane. However, the temperature is significantly lower than what we have estimated above. This is certainly due to the fact that \citet{Fre05} did not include He\,{\sc ii} photospheric absorption features in their analysis. The projected rotational velocity of the star inferred by \citet{Fre05} is close to the median value (236\,km\,s$^{-1}$) of the published $v\,\sin{i}$ values \citep{Con77,How89,Pen96,How97,Fre05,Rau15}.

\subsection{Short-term variability}
In Sect.\,\ref{varphotom}, we discussed a weak photometric modulation at a frequency of $\nu_1 = 0.335$\,d$^{-1}$ in the {\it Hipparcos} and  ASAS-3 data. The two usual suspects for explaining such a variability are rotational modulation of a spotted star and (non-radial) pulsations \citep{Baa94}. Despite many years of investigation of photometric and spectroscopic variability of Be stars, there is still no consensus as to whether the short-term variability stems from rotational modulation or (multiperiodic) non-radial pulsations \cite[e.g.][]{Sem18,Har19}. 

Let us start by considering the hypothesis of the rotational modulation. From the above median value of $v\,\sin{i} = 236$\,km\,s$^{-1}$, we can estimate upper limits on the rotation period of 2.71 and 3.11\,days for photometric radii of 12.6 and 14.5\,R$_{\odot}$ respectively. If we assume an inclination of the rotation axis of $35^{\circ}$ \citep{Fre05}, the actual rotation periods become 1.57 and 1.80\,days, which are both significantly shorter than the photometric 2.984\,days period. Rotational modulation thus seems rather unlikely to explain the observed periodicity. 

If we interpret instead the $\nu_1$ frequency as the signature of non-radial pulsations, its value would suggest pulsations similar to those of slowly pulsating B-stars. The overall stellar properties inferred hereabove place HD~60\,848 in a temperature and luminosity range where much faster $\beta$~Cep - like pulsations would be traditionally expected. However, the instability strip calculations of \citet{Godart} indicate that some stars in this part of the Hertzsprung-Russell diagram can display both high- and low-frequency modes. In this context, we recall the work of \citet{Boy07} who analysed 62 spectra of HD~60\,848 collected in December 2000. Applying a cross-correlation method, they measured the RV of the H$\alpha$ emission and found rapid RV variations with putative periods of $3.51 \pm 0.03$\,hr and $3.74 \pm 0.03$\,hr, corresponding to frequencies of $6.84 \pm 0.06$\,d$^{-1}$ and $6.42 \pm 0.05$\,d$^{-1}$ respectively. They attributed these variations to non-radial pulsations affecting the underlying photospheric absorption profile. Similar conclusions were reached by \citet{McS07} based on a set of blue spectra of the star which did not reveal evidence for binarity, but displayed short-term RV variations of the absorption lines possibly due to pulsations. Such changes could explain the RV dispersion we found in our data (Sect.\,\ref{specvar}). Only the lower of the two frequencies found by \citet{Boy07} would be marginally consistent with $\nu_1 + 6$\,d$^{-1}$. As stressed in Sect.\,\ref{varphotom}, the currently available photometric data do not properly sample such short periods, preventing us from reaching a firm conclusion as to what frequency is the right one. Therefore, whilst pulsations appear as an attractive explanation, the very nature of these pulsations (SPB versus $\beta$~Cep) remains currently an open question.   

\subsection{Recurrent disk cycles}
Our dataset revealed the existence of recurrent, but irregular cycles in the strength of the H$\alpha$ emission line. Before we interpret these variations, we briefly consider the observational evidence for the presence or absence of a circumstellar disk in HD~60\,848. Using polarization measurements through narrow-band filters, \citet{Coy76} detected a decrease of the linear polarization across the H$\beta$ line by 18\% relative to the polarization in the continuum. This depolarization effect is expected for a rotating circumstellar disk \citep[e.g.][]{Poe77}. This finding contrasts with the non-detection of a depolarization across the H$\alpha$ line by \citet{Vin09} who thus questionned the mere existence of a circumstellar disk around the star. However, a weak depolarization such as the one measured by \citet{Coy76} would remain undetected in the \citet{Vin09} data considering their large error bars. The low level of polarization (and depolarization) in the spectra of Oe stars compared to some Be stars could stem from the existence of a rather strong stellar wind at stellar latitudes above the disk, especially for a star seen nearly pole-on, such as HD~60\,848. This wind contributes a nearly-spherical population of free electrons that scatter photons from the star and from the disk thus reducing the overall level of polarization. We thus consider that the lack of detection of a depolarization effect in the H$\alpha$ line by \citet{Vin09} is not an argument against the line arising from a Be-like circumstellar decretion disk. 

Recurrent, but non-periodic outbursts have been observed in the optical lightcurves of several Be stars \citep[e.g.][]{Car03,Gho18,Lab18}. Among the best-studied objects, the B2.5\,Ve star $\omega$~CMa displayed four cycles of disk formation and partial dissipation over a duration of 34\,yr as seen in optical photometry \citep{Gho18}. The timescales of disk growth and decay were found to vary from cycle to cycle \citep{Gho18}.

Most variability studies of Be disks rely on time series of optical photometry. Different stars display different relations between the H\,{\sc i} line emission strength and optical brightness. For instance, $\gamma$~Cas displays a positive correlation between H$\alpha$ strength and $V$-band brightness \citep{Pol14}, whereas negative correlations are seen for other objects \citep[e.g.\ HD~45\,314,][]{Rau18}. \citet{Har83} suspected that positive or negative correlations between optical brightness and H\,{\sc i} line emission strength reflect a geometrical effect ruled by the inclination of the circumstellar envelope with respect to the observer's line of sight. \citet{Sig13} theoretically investigated this suggestion and found that the existence of a positive or inverse correlation between EW(H$\alpha$) and $V$ depends on the scale height of the disk and the importance of gravity darkening of the star. \citet{Rim18} defined three types of Be lightcurve behaviour depending on the inclination of the disk. For $i \leq 70^{\circ}$ (near pole-on orientation), an outburst leads to an increase of the brightness in $V$ (i.e.\ positive correlation), whereas for $i \geq 85^{\circ}$ (shell stars), a disk outburst will produce dips in the lightcurve (i.e.\ negative correlation). In between these two regimes, the lightcurve displays a complex behaviour depending on the balance between extra reflection and absorption of photospheric light.

The most-likely inclination of the disk puts HD~60\,848 into the first category as defined by \citet{Rim18}. Yet, our results show no clear correlation between optical brightness and EW(H$\alpha$). This situation most likely reflects the fact that the various quantities trace different parts of the disk. Indeed, optical emission from a disk most likely arises from the inner 1 -- 2\,R$_*$, whereas H$\alpha$ emission for instance arises over a significantly wider range of distances in the disk \citep[typically out to 15\,R$_*$, see][]{Lab18}. During disk dissipation, the equivalent width of an emission line arising over a wide range in radii may thus still increase whilst the optical brightness is already decaying. In fact, if the disk dissipation proceeds from inside out, the inner disk regions are removed first, leading to a rapid decline of the disk's optical continuum emission. The absolute line flux also decreases, but its relative decrease is slower than that of the disk's optical continuum, resulting in a phase of relative increase of the emission line strength \citep{Sig13,Lab18}.

\citet{Vie17} used the {\it pnrc} parameters of \citet{Fre05} in conjunction with the viscous decretion disk model and IR flux measurements to infer the disk density scale and its dependence with radius for a large set of Be stars\footnote{In their model, the number density of the disk is given by $n(r,z) = n_0\,\left(\frac{r}{R_*}\right)^{-\alpha}\,\exp{\left[-\frac{1}{2}\left(\frac{z}{H(r)}\right)^2\right]}$ with $H(r)$ the disk scale height, $r$ the radial coordinate and $z$ the elevation above the disk plane.}. These authors found that in optical photometry the time-scale for disk growth is generally shorter than the time-scale for disk dissipation \citep[see also][]{Lab18}. Using the same kind of models, \citet{Rim18} showed that disks that build-up over a longer duration accumulate a mass reservoir in the outer parts of the disk and their dissipation would take longer. This description faces some problems in the case of HD~60\,848. Whilst the existing photometry is not fully conclusive regarding the rise and fall times, it suggests that HD~60\,848 was generally brighter between 2014 and 2018, which would hint at a relatively stable build-up. Until 2016, the increase of the H$\alpha$ emission strength was rather slow, suggesting indeed accumulation of a mass reservoir in the outer disk parts. However, the decay of the H$\alpha$ emission in 2001, 2009, and 2018 -- 2019 occurred much faster than the disk build-up. This fast decay might reflect efficient ablation of the disk by the strong radiation field of the Oe star. Indeed, \citet{Kee16} showed that the strong radiative force of O-stars significantly shortens the disk destruction time compared to later-type Be stars where this process is much less efficient.   

In the specific case of HD~60\,848, \citet{Vie17} used {\it WISE} observations taken in 2010 (January to November) to infer $\alpha = 3.0 \pm 0.2$ and a steady-state disk decretion rate of $8\,10^{-12}$\,M$_{\odot}$\,yr$^{-1}$ for a viscosity parameter unity. The $\alpha$ value is right at the border between steady-state and dissipating disks as defined by \citet{Vie17}. However, according to our H$\alpha$ data, the disk was in a low emission state at this epoch, following the rapid decline that took place in 2009. It thus seems that some of the parameters of the circumstellar disk of HD~60\,848 follow a different behaviour than for disks around cooler Be stars. 

\section{Conclusion} \label{Conclusion}
Optical spectroscopy and photometry of the O9e star HD~60\,848 revealed cyclic, but irregular, variations of the circumstellar material. Several built-up and decay events were detected, but they are not ruled by a strict periodicity. No significant periodicity was found in the RVs or characteristics of the emission line profiles. There exists no direct correlation between the photometric variability and the changes in the strength of the H$\alpha$ emission, and we found a time delay between the variations of the EWs of H$\alpha$ and H$\beta$ on the one hand, and He\,{\sc i} $\lambda$\,5876 on the other hand. We suggest that this behaviour stems from a two-step build-up phase of the disk, where the growing disk first expands radially and then increases its density. In contrast, the optically thin H\,{\sc i} Paschen emission lines follow the photometric variations more closely. Whilst most of the properties of the circumstellar material are consistent with those of Keplerian decretion disks, some aspects most probably reflect the earlier spectral type of the star. The most obvious example is certainly the very fast decay of the H$\alpha$ emission that we have observed which suggest that the wind and the radiation field of the star play a significant role in the disk erosion. Finally, we found no significant impact of the disk state on the X-ray emission of the star. More specifically, we found no hardening or brightening of the X-ray emission when the star was in the high emission state. Instead, the X-ray emission of this star is entirely compatible with the emission arising from a radiatively-driven wind of a normal O-type star. 

\section*{Acknowledgements}
We are grateful to Dr Belinda Wilkes, Director of the {\it Chandra} X-ray Center for granting us a ToO observation on the Director's Discretionary Time. The Li\`ege team thanks the European Space Agency (ESA) and the Belgian Federal Science Policy Office (BELSPO) for their support in the framework of the PRODEX Programme (contract XMaS and HERMeS). We acknowledge with thanks the variable star observations from the AAVSO International Database contributed by observers worldwide and used in this research as well as the amateur spectroscopists who contributed data to the BeSS database. The ADS, CDS and SIMBAD databases were used in this work.

\end{document}